\documentclass[useAMS,usenatbib,twocolumn]{mn2e}

\usepackage[dvipsnames]{color}
\usepackage{graphicx}
\usepackage{placeins}
\usepackage{flushend}
\usepackage{cuted}
\usepackage{widetext}
\usepackage{bm}
\usepackage{natbib}
\bibpunct{(}{)}{;}{a}{}{,}
\usepackage{amssymb}
\usepackage{amsmath}
\usepackage{times}
\usepackage[]{eufrak}
\usepackage{textcomp}
\usepackage{aecompl}
\usepackage[T1]{fontenc}

\newcommand{\ie}{\textit{i.e.},~}
\newcommand{\cf}{\textit{cf.},~}

\newcommand{\be}{\begin{equation}}
\newcommand{\ee}{\end{equation}}
\newcommand{\bdm}{\begin{displaymath}}
\newcommand{\edm}{\end{displaymath}}
\newcommand{\bea}{\begin{eqnarray}}
\newcommand{\eea}{\end{eqnarray}}

\title[Subpulse velocity in SCLF]
{Explaining the subpulse drift velocity of pulsar magnetosphere
within the space-charge limited flow model}

\author[V.~S.~Morozova, B.~J.~Ahmedov and O.~Zanotti]
        {Viktoriya~S.~Morozova$^{(1,\;2)}$,
          Bobomurat~J.~Ahmedov$^{(3,\;4,\;5)}$  and
        Olindo~Zanotti$^{(6)}$
                                                                \\
  $^{(1)}$Max-Planck-Institut f$\ddot{u}$r
        Gravitationsphysik, Albert-Einstein-Institut,
        Am M{\"u}hlenberg 1, 14476 Potsdam, Germany \\
        $^{(2)}$Theoretical Astrophysics, California Institute of Technology,
        Pasadena, CA 91125, USA \\
 $^{(3)}$Institute of Nuclear Physics,
        Ulughbek, Tashkent 100214, Uzbekistan                   \\
        $^{(4)}$Ulugh Beg Astronomical Institute,
        Astronomicheskaya 33, Tashkent 100052, Uzbekistan       \\
        $^{(5)}$The Abdus Salam International Centre
for Theoretical Physics, 34151 Trieste, Italy\\
$^{(6)}$ Laboratory of Applied Mathematics,
University of Trento, Via Mesiano 77, 38123 Trento, Italy \\
\\}
\pagerange{\pageref{firstpage}--\pageref{lastpage}}

\pubyear{2014}

\begin{document}

\maketitle

\label{firstpage}

\begin{abstract}

We try to explain the subpulse drift
phenomena adopting the space-charge limited
flow (SCLF) model and comparing the plasma drift velocity
in the inner region of pulsar magnetospheres with the
observed velocity of drifting subpulses. We apply
the approach described in a recent paper of
\cite{lt12}, where it was shown that the standard estimation of
the subpulse drift velocity through the total value of
the scalar potential drop in the inner gap
gives inaccurate results, while the
exact expression relating the drift velocity to the
gradient of the scalar potential should be used
instead.
After considering a selected sample of sources taken from
the catalog of \cite{wes06} with coherently drifting
subpulses and reasonably known observing geometry,
we show that their
subpulse drift velocities
would correspond to the drift of
the plasma located very close or above the pair formation
front. Moreover, a detailed analysis of
PSR B0826-34 and PSR B0818-41 reveals that
the variation of the subpulse separation with the pulse
longitude can be successfully explained
by the dependence of the plasma drift velocity on the
angular coordinates.

\end{abstract}


\section{Introduction}

The research on pulsar magnetospheres started 45 years ago from
the pioneering paper of \cite{gj69}, where it was shown that the
magnetic field, together with the fast rotation of the pulsar,
generates strong electric fields tending to pull out charged
particles from the surface of the neutron star and to form the
plasma magnetosphere. However, the very next arising question -
how many particles will actually leave the surface of the star
under the action of this force - is still a subject of scientific
debate. In fact, the magnetic field which generates the pulling out
electric field also leads to the substantial increase in the
cohesive energy of the surface charged particles (positive ions in
larger degree than electrons), making the outer layer of the star
very dense and strongly bound. \cite{ml07} showed that for each
magnetic field intensity there exists a critical surface
temperature, above which the particles are able to freely escape
from the surface of the star. The magnetic field of pulsars is
typically inferred from observations under the assumption that it
is strictly dipolar, thus providing $B_d = 2\times
10^{12}\sqrt{P\dot{P}\times10^{15}}\mathrm{G}$ (here $P$ is
the period and $\dot{P}$ is the period derivative of the pulsar).
Using this formula, it turns out that the majority of pulsars
satisfy the condition for the free particle outflow. However, in
many works, starting from \cite{rs75} it was suggested that the
magnetic field close to the surface of the pulsar
should have a multipole structure, with the surface magnetic field
several orders of magnitude larger than the estimated $B_d$. This
idea is supported by X-ray observations
\citep{zsp05,kpg06,pkwg09}, and several studies have been
performed searching for a mechanism of generation and maintenance
of such small-scale strong magnetic fields [see \cite{ggm13} and
references therein].

The amount of charged particles extracted from the surface of the
star by the rotationally induced electric field is a key aspect of
any pulsar magnetosphere model. The model of \cite{rs75}, for
example, assumes that no particle leaves the pulsar surface and
there is a vacuum gap formed above the star with a huge difference
in the scalar potential between the bottom and the top, \ie $\sim
10^{12} \mathrm{V}$. According to this idea, the gap will be
periodically discharged and rebuilt, making it intrinsically
"non-stationary". On the contrary, in the "stationary"
space-charge limited flow (SCLF) model of \cite{as79} there is a
free flow of charged particles from the pulsar surface. The
partially screened gap model of \cite{gmg03} combines some
features of both the previous models and is based on  the
assumption that the vacuum gap discharge will lead to the back
flow of charged particles bombarding and heating the surface of
the star, causing the thermal ejection of ions from the surface
and partially screening the original accelerating electric field.

The choice among different models of pulsar
magnetosphere should be made by accurate comparison of
their predictions with the results of observations. A
very interesting phenomena serving as a diagnostic tool
is the subpulse drift. Pulsar radio emission comes to us
in the form of pulses, which may look very different
among each other and typically consist of individual
subpulses.
%or disappear on a time interval from several to hundreds
%of pulsar periods.
Although the average pulse profile is very stable and represents a
unique \emph{fingerprint} of each pulsar, a range of successive
pulses plotted on top of each other in a so-called pulse stack,
quite often shows an organized phase shift of the subpulses
forming drift bands. This phenomenon was reported for the first
time in \cite{dc68}, while the first systematic analysis of
``drifting'' pulsars dates back to \citet{b81} and \citet{r86}. So
far, the largest statistical study of the phenomenon has been
presented in the works of \cite{wes06} and \cite{wse07}, who
considered a sample of 187 pulsars, $55\%$ of which show drifting
subpulses. Usually the subpulse drift bands are characterized by
the horizontal separation between them in subpulse longitude,
$P_2$, and the vertical separation in pulse periods, $P_3$. The
subpulse behavior of the individual pulsars may be rather complex
and demonstrate smooth or abrupt change of drift direction, phase
steps, longitude and frequency dependence of the separation $P_2$,
or even presence of subpulses drifting in the opposite directions
at the same time. {For the graphical representation of the periods
$P_2$ and $P_3$ we refer the reader to Fig.~1 of \cite{wes06},
where one can find several examples of the different subpulse
behavior of individual sources.}

The vacuum gap model, and especially the partially
screened gap model, has been widely used to explain the
subpulse drift phenomena
\citep{gs00,mgp00,gmg03,ggks04,bggs07,getal08}, while
for a long time
the SCLF model has been regarded unable to account for it.
However, recent analytical \citep{lt12} and
numerical \citep{t10b,ta13} progresses
have shown that the door can be left
open even for the SCLF model.
The main goal of this paper is to explain the subpulse
drift velocity in the
framework of the relativistic SCLF 
model, without addressing the
question of the generation mechanism of the plasma
features responsible for the appearance of the subpulses,
while trying to compare the results with the available
observational data.

The plan of the paper is the following. In Sect.~\ref{oldmodels} we
give a brief review of the models used so far to explain the
subpulse phenomena and motivate our choice to concentrate on the
SCLF model. In Sect.~\ref{velocitySCLF} we present the basic
equations to explain the subpulse drift velocity in the framework
of the SCLF model. In Sect.~\ref{alt} we consider a set of pulsars
from the catalog of \cite{wes06} with coherently drifting
subpulses, and try to deduce in which regions of the pulsar
magnetosphere the SCLF model would predict the plasma with the
observed velocities. In Sect.~\ref{individualpulsars} we focus on
two specific sources, PSR B0826-34 and PSR B0818-41, trying to
account for their phenomenology. Finally, Sect.~\ref{conclusions}
is devoted to the summary of the results obtained and to the
conclusions.

%%%%%%%%%%%%%%%%%%%%%%%%%%%%%%%%%%%%%%%%%%%%%%%%%
\section{A brief survey of existing models}
\label{oldmodels}
%%%%%%%%%%%%%%%%%%%%%%%%%%%%%%%%%%%%%%%%%%%%%%%%%

The first theoretical explanation of the subpulses was provided by
\cite{rs75}, who associated the subpulses to the spark discharges
of the vacuum gap above the pulsar surface. In its original form,
the model applied to the pulsars with anti-parallel angular
velocity $\vec{\Omega}$ and magnetic moment $\vec{\mu}$, and
assumed that the charged particles (positive ions) are tightly
bound to the surface of the star and cannot be pulled out by the
rotationally induced electric field. This requirement leads to the
formation of a vacuum gap in the region where the magnetic field
lines are open and with a potential drop between the top and the
bottom of the order of $10^{12} \mathrm{V}$ for typical pulsar
parameters. Due to the presence of strong curved pulsar magnetic
fields, the gap will be unstable and periodically discharged by
the photon induced pair creation process. The discharges will
build up plasma columns, which are subject to the $\vec{E}\times
\vec{B}$ drift in the electromagnetic field of the magnetosphere.
\cite{rs75} showed that, unless the potential drop of the gap is
completely screened, the plasma columns will not exactly corotate
with the star but always lag behind the rotation of the star, and
this is responsible for the visual drift of the subpulses along
the pulse longitude. The sparks are assumed to form rings and the
so-called tertiary periodicity $P_4$ is the time needed for the
spark carousel to make one full rotation around the magnetic axis.
Although in this model the subpulses cannot outrun the rotation of
the star, due to the effect of aliasing \citep{gmg03,ggks04},  the
apparent velocity of the subpulses may be both positive (from
earlier to later longitudes) and negative (from later to earlier
longitudes).

\citet{rs75} estimated  the subpulse drift velocity
to be proportional to the full potential drop across the
gap, resulting in excessively large values of the drift
velocity compared with the observed ones. Later on,
\citet{gs00}, \citet{mgp00}, \citet{gmg03} generalized
this model to account for arbitrary inclination angles
$\chi$ between $\vec{\Omega}$ and $\vec{\mu}$, and
modified it to allow for the partial outflow of ions and
electrons from the surface of the star, forming partially
screened gap instead of the pure vacuum.  It was argued
that the favorable conditions for the spark discharge
persist even if the original vacuum gap is screened up to
$95\%$ or more, making
the velocity of drifting subpulses
consistent with the observed values.
However, the partially screened gap model requires
surface values of the magnetic field of the order
of $10^{14} \mathrm{G}$, much larger than those  deduced
when the magnetic field is dipolar,\footnote{
There are mechanisms which can provide long
living small scale magnetic field of the required
strength at the surface of the pulsar \citep{ggm13}.
}
i.e. $\sim 10^{12} \mathrm{G}$.
The partially screened gap model has been  used in a
number of works to describe the subpulses of specific
pulsars as well as their X-ray emission
\citep{ggks04,bggs07,getal08} and
it has typically revealed a strong predictive power.

In \cite{cr04,rc08,rd11}, the drifting subpulses are instead
explained by non-radial oscillations of the surface of
the star. This model gives a very natural explanation to
the subpulse phase shift, relating it to the intersection
of the observer's line of sight with the nodal line. The
empirical model of \cite{w03} relates the formation of
drifting subpulses to the interaction between electron
and positron beams traveling up and down between the
inner and the outer gaps of the pulsar
magnetosphere.

\cite{kmm91} and later \cite{gmml05} proposed a model where the
subpulses are formed due to the modulation of the emission region
by large-scale ``drift waves'', generated by oppositely directed
curvature drifts of electrons and positrons. Finally, \cite{fkk06}
proposed a model to explain drifting subpulses that is  based on
the diocotron instability in  the pair plasma on the open field
lines. Detailed description of all these models may be found in
the review of \cite{k09} and references therein.

% - read and cite Jones_2011

Recently \cite{lt12} have shown that the order of magnitude
estimation of the subpulse drift velocity used in \cite{rs75}, and
subsequent works, can be replaced by a more precise expression,
relating the velocity to the radial derivative of the potential
instead of the absolute value of the potential drop. This simply
comes from the fact that the subpulse drift velocity is determined
by the $\mathbf{E}\times \mathbf{B}$ drift of the distinct plasma
features in the magnetosphere, while the electric field
responsible for this drift is given by the gradient of the scalar
potential. The final expression for the subpulse drift velocity in
degrees per period is~\citep{lt12}
\begin{equation}
\label{lt}
\omega_D = \frac{180^\circ}{\xi} \frac{d\tilde{V}}{d\xi}\,,
\end{equation}
where $\tilde{V}\equiv V/\Delta V_{\rm vac}$ is the
scalar potential normalized to the potential drop $\Delta
V_{\rm vac} = \Omega B r_{pc}^2 /2c$ between the rotation
axis and the boundary of the (small) polar cap, $r_{pc}$ is the
radius of the polar cap, $B$ is the value of the magnetic
field (which \cite{lt12} assumed to be constant across
the  polar cap), $\xi\equiv \theta/\Theta$ is the
polar angle normalized to the colatitude of the polar cap
boundary $\Theta$, $c$ is the speed of light. As
emphasized in \cite{lt12}, this velocity, being the true
plasma drift velocity in the open magnetic field line
region, should introduce periodic modulation to the
observed radio emission from pulsars, independently on
the particular emission generation mechanism.
We stress that this periodic modulation (which is exactly the tertiary
periodicity, mentioned before) can be found even in the spectra
of those pulsars which don't reveal regularly drifting
subpulses~\citep{getal08,gmz07}.
The same is seen for many pulsars listed in \cite{wes06} and \cite{wse07}.

The main reason for which the SCLF model has been so far
regarded unable to explain the
subpulse drift phenomena is that it has not
a prescribed mechanism for the formation of
spark-like features, which are invoked to explain
the subpulses in the vacuum and in the partially screened gap
models.
However, recent progress in numerical simulations of
pulsar magnetosphere have shed new light
on this subject. One of the main tendencies in
the numerical studies of pulsar magnetosphere is to
consider it as a global object, with the different
regions closely interlinked and interdependent. For example, self-consistent
simulations of pair cascades in the polar cap region
\citep{t10b,ta13} show that the cascade behavior is
mostly determined by the global magnetospheric current
density and that periods of plasma generation are
interleaved with quiet periods,
%when the electric
%field is screened and no pairs are produced,
both for the vacuum gap and for the SCLF regimes. Based on that,
\cite{lt12} proposed the idea that the distinct emitting features
in the inner magnetosphere, responsible for the appearance of the
subpulses, may be caused by global current filaments, similar to
those observed in auroras.

These arguments motivated us to adopt the approach of \cite{lt12}
for studying the possible subpulse behaviour within the SCLF
model. In the rest of the paper we do not consider the problem of
the subpulse generation. Rather, we concentrate on the velocity
given by Eq.~(\ref{lt}), calculated from the analytical
expressions for the scalar potential in the 
relativistic SCLF model,
 which are available in the literature (see \cite{mt92,hm98,hm01,hm02}).
The main issue that we try to address is whether it is possible,
in the framework of the SCLF model, to explain the observed
subpulse drift velocities, and if so, to infer in which part of
the magnetosphere they should be produced. In the second part of
the paper we will instead apply our arguments to two specific
sources.

%%%%%%%%%%%%%%%%%%%%%%%%%%%%%%%%%%%%%%%%%%%%%%%%%
\section{Subpulse drift velocity in the framework of the SCLF model}
\label{velocitySCLF}
%%%%%%%%%%%%%%%%%%%%%%%%%%%%%%%%%%%%%%%%%%%%%%%%%

\cite{saf78} and \cite{as79} were the first to show analytic
solutions for the scalar potential in the vicinity of the pulsar
polar cap and in the framework of the  SCLF model. In their
analysis, the accelerating electric field parallel to the magnetic
field of the pulsar is due to the curvature of magnetic
field lines and to the inertia of particles. Later, \cite{mt92}
have shown that, due to the effect of dragging of inertial
frames in general relativity, it is possible to obtain
accelerating electric fields which are two orders of magnitude
larger than those normally expected. This approach has further been
elaborated in \cite{hm98,hm01,hm02}. For convenience, in this
subsection we present the main results found by \cite{mt92} as
well as the expressions for the subpulse drift velocity that we
obtained, \ie using Eq.~(\ref{lt}).

In general relativity the dipole-like magnetic field in the
exterior spacetime close to the surface of a slowly rotating
neutron star described by the metric
\begin{eqnarray}
\label{metric}
&& ds^2=-N^2c^2dt^2+N^{-2}dr^2 \nonumber \\ && \qquad\qquad
+r^2d\theta^2+r^2\sin^2\theta d\phi^2-
2\omega r^2\sin^2\theta dt d\phi\
\end{eqnarray}
is given by the expressions
\begin{eqnarray}
\label{Br}
&&\hat{B}_r = B_0\frac{f(\bar{r})}{f(1)}\bar{r}^{-3}\cos\theta \,,  \\
\label{Btheta}
&&\hat{B}_{\theta} = \frac{1}{2} B_0 N \left[ -2\frac{f(\bar{r})}{f(1)} +\frac{3}{(1-\varepsilon/\bar{r})f(1)} \right] \bar{r}^{-3}\sin\theta \,.
\end{eqnarray}
Here the spherical coordinates $(r,\theta,\phi)$ are used with the
polar axis oriented along the magnetic moment of the pulsar,
$\bar{r} = r/R$, $R$ is the radius of the neutron star, $B_0 =
2\mu/R^3$ is the value of the magnetic field at the pole, $ N  =
(1-2GM/rc^2)^{1/2}$ is the lapse function of the metric, $G$ is the
gravitational constant, $M$ is the mass of the star, $\omega$
is the frequency of dragging of inertial frames,
$\varepsilon=2GM/Rc^2$ is the compactness parameter, while the
function $f(\bar{r})$ is given by

\begin{equation}
\label{f}
f(\bar{r}) = -3\left(\frac{\bar{r}}{\varepsilon}\right)^3 \left[\ln\left(1-\frac{\varepsilon}{\bar{r}}\right) + \frac{\varepsilon}{\bar{r}} \left(1 + \frac{\varepsilon}{2\bar{r}}\right) \right]\,.
\end{equation}
The polar angle of the last open magnetic field line $\Theta$ is equal to
\begin{equation}
\Theta\cong\sin^{-1}\left\{\left[\bar{r}\frac{f(1)}{f(\bar{r})}\right]^{1/2}\sin\Theta_0\right\}\,,
\end{equation}
where
\begin{equation}
\label{theta0}
\Theta_0 = \sin^{-1}\left(\frac{R\Omega}{c f(1)}\right)^{1/2}
\end{equation}
is the polar angle of the last open magnetic field line at the surface of the star.

The scalar potential $\Phi$ in the polar cap region of
the inner pulsar magnetosphere is obtained from the
solution of the equation $\Delta\Phi =
-4\pi(\rho-\rho_{GJ})$, where $\rho_{GJ}$ is the
Goldreich--Julian charge density and $\rho$ is the actual
charge density in the open field line region of the
magnetosphere. The boundary conditions
of this Poisson equation are
(i) $\Phi=0$ at the surface of the star and along the last open
magnetic field lines, and (ii) $E_{\parallel}=0$ at large
distances from the star.
 Very close to the surface of the star ($\bar{r}-1<<1$) the solution reads

\begin{widetext}
\begin{eqnarray}
\label{Philow}
&&\Phi_{\rm low} = 12 \frac{\Phi_0}{\bar{r}} \sqrt{1-\varepsilon} \kappa \Theta_0^3 \cos\chi \sum_{i=1}^{\infty}\left[ \exp{\left(\frac{k_i(1-\bar{r})}{\Theta_0\sqrt{1-\varepsilon}}\right)} - 1 + \frac{k_i(\bar{r} -1)}{\Theta_0\sqrt{1-\varepsilon}} \right] \frac{J_0(k_i\xi)}{k_i^4 J_1(k_i)} \nonumber \\  &&\qquad\qquad
+ 6\frac{\Phi_0}{\bar{r}} \sqrt{1-\varepsilon} \Theta_0^4 H(1) \delta(1) \sin\chi \cos\phi \sum_{i=1}^{\infty}\left[ \exp{\left(\frac{\tilde{k}_i(1-\bar{r})}{\Theta_0\sqrt{1-\varepsilon}}\right)} - 1 + \frac{\tilde{k}_i(\bar{r} -1)}{\Theta_0\sqrt{1-\varepsilon}} \right] \frac{J_1(\tilde{k}_i\xi)}{\tilde{k}_i^4 J_2(\tilde{k}_i)}\,,
\end{eqnarray}
while at the distances $\Theta_0<<\bar{r} - 1<< c/\Omega R$ one gets
\begin{equation}
\label{Phihigh}
\Phi_{\rm high}=\frac{1}{2} \Phi_0 \kappa \Theta_0^2 \left(1-\frac{1}{\bar{r}^3}\right) \left(1-\xi^2\right) \cos\chi + \frac{3}{8}\Phi_0 \Theta_0^3 H(1) \left(\frac{\Theta(\bar{r}) H(\bar{r})}{\Theta_0 H(1)}-1\right) \xi(1-\xi^2) \sin\chi \cos\phi\,.
\end{equation}
\end{widetext}
Here $\Phi_0=\Omega B_0 R^2/c$,
$\kappa\equiv\varepsilon\beta$, while $\beta=I/I_0$ is the
stellar moment of inertia in units of $I_0=MR^2$. The parameter $\xi =
\theta / \Theta$
changes from 0 to 1 inside the polar
cap region, $J_m$ is the Bessel function of order $m$,
$k_i$ and $\tilde{k}_i$ are the positive zeroes of the
Bessel functions $J_0$ and $J_1$, arranged in ascending
order. Moreover,
\begin{equation}
H(\bar{r}) = \frac{1}{\bar{r}}\left(\varepsilon - \frac{\kappa}{\bar{r}^2}\right) + \left(1-\frac{3}{2}\frac{\varepsilon}{\bar{r}}+\frac{1}{2}\frac{\kappa}{\bar{r}^3}\right)\left[f(\bar{r})\left(1-\frac{\varepsilon}{\bar{r}}\right)\right]^{-1}\,,
\end{equation}
and $\delta(\bar{r}) = d \ln[\Theta(\bar{r}) H(\bar{r})]/d\bar{r}$.

These results allow one to find the plasma drift velocity
\begin{equation}
\vec{v}_D=c\frac{\vec{E}\times\vec{B}}{B^2}
\end{equation}
in the polar cap region of the magnetosphere with the electric
field $\vec{E}=-\nabla\Phi$ and the magnetic field
(\ref{Br})--(\ref{Btheta}). One can easily show that the largest
contribution to the azimuthal drift in the corotating frame of the
star is due to the term
$-({cE_{\theta}B_r}/{B^2})\hat{\phi}$, which, after proper
transformations (see subsection 2.2 of \cite{lt12} for the
details), leads to the subpulse drift velocity in degrees per
period as given by (\ref{lt}). The final expressions for the drift
velocity, obtained from (\ref{Philow}) and (\ref{Phihigh}) using
$J_0^\prime (x)=-J_1 (x)$ and $J_1^\prime (x)=(J_0(x)-J_2(x))/2$
look like

\begin{widetext}
\begin{eqnarray}
\label{omegaDlow}
&&\omega_{D\ \rm low} = \frac{180^\circ}{\xi} \frac{12\sqrt{1-\varepsilon}\Theta_0}{\bar{r}} \Bigg\{ -2\kappa\cos\chi \sum_{i=1}^{\infty}\left[ \exp{\left(\frac{k_i(1-\bar{r})}{\Theta_0\sqrt{1-\varepsilon}}\right)} - 1 + \frac{k_i(\bar{r} -1)}{\Theta_0\sqrt{1-\varepsilon}} \right] \frac{J_1(k_i\xi)}{k_i^3 J_1(k_i)} \nonumber \\ &&\qquad\qquad
 +\Theta_0 H(1) \delta (1) \sin\chi \cos\phi \sum_{i=1}^{\infty}\left[ \exp{\left(\frac{\tilde{k}_i(1-\bar{r})}{\Theta_0\sqrt{1-\varepsilon}}\right)} - 1 + \frac{\tilde{k}_i(\bar{r} -1)}{\Theta_0\sqrt{1-\varepsilon}} \right] \frac{J_0(\tilde{k}_i\xi) - J_2(\tilde{k}_i\xi)}{2 \tilde{k}_i^3 J_2(\tilde{k}_i)} \Bigg\}
\end{eqnarray}
and
\begin{equation}
\label{omegaDhigh}
\omega_{D\ \rm high} = \frac{180^\circ}{\xi}\left[ -2\xi\kappa\left( 1-\frac{1}{\bar{r}^3} \right) \cos\chi  + (1-3\xi^3) \frac{3}{4} \Theta_0 H(1) \left(\frac{\Theta(\bar{r}) H(\bar{r})}{\Theta_0 H(1)} -1\right) \sin\chi \cos\phi \right]\ .
\end{equation}
\end{widetext}

For inclination angles $\chi<90^\circ$ (except for the
almost orthogonal pulsars) the scalar potential in the
polar cap region is positive, has a maximum close to the
magnetic axis and goes to zero at the last open magnetic
field lines, so that the value of $\omega_D$ is negative
almost everywhere. From the point of view of observations
it means that the SCLF model predicts negative drift
(from larger to smaller longitudes) in case of the outer
line-of-sight geometry and positive drift in case of the
inner line-of-sight geometry.
In the rest of our work we claim that the velocities
(\ref{omegaDlow}) and (\ref{omegaDhigh}) represent the true drift
velocities of whatever features are responsible for the
subpulses in a specific portion of the magnetosphere [see
also \cite{lt12}].
Moreover, the expressions  (\ref{omegaDlow}) and
(\ref{omegaDhigh}) predict the longitude dependent (not
constant) apparent drift velocity of the subpulses along
any observer's line of sight, unless the inclination
angle of the pulsar is exactly zero and the line of sight is
exactly concentric with the magnetic field
axis. In section \ref{individualpulsars} we will use this fact to explain
the longitude dependence of the subpulse separation in case of two individual pulsars.

\begin{table*}%[ht]
%\begin{widetext}
\centering
\caption{Coherent drifters from the catalog of
  \citet{wes06} for which the inclination angle $\chi$ is
  known from previous studies.
\newline $^a$ Value taken from \citet{r93}. \newline $^b$ Value taken from \citet{lm88}.}
\begin{tabular}{c c c c c c c c c}
\hline\hline
Pulsar name & $P$ (s) & $\dot{P}$ & $P_2$ $(^\circ)$ [21 cm] & $P_3$ ($P$) [21 cm] & $P_2$ $(^\circ)$ [92 cm] & $P_3$ ($P$) [92 cm] &  $\chi$ $(^\circ)$ & $\beta$ $(^\circ)$    \\ [0.5ex] % inserts table %heading
\hline
B0148-06 & 1.4647 & $4.4 \times 10^{-16}$ & $-12.5\ ^{+0.4}_{-1.9}$ & $14.2\pm 0.2$ & $-14\ ^{+0.6}_{-0.5}$ & $14.4\pm 0.1$ & 14.5 & $1.9^{a},\ 2.1^{b}$   \\

B0149-16 & 0.8327 & $1.3 \times 10^{-15}$ & $-9     \ ^{+12}_{-1}    $ & $5.8  \pm 0.5$ & $-13\ ^{+2}_{-4}$ & $5.7\pm 0.2$ & 84    & $1.9^{a},\ 1.9^{b}$   \\

B0320-39 & 3.0321 & $6.4 \times 10^{-16}$ & $-18   \ ^{+5}_{-3}      $ & $8.4  \pm 0.1$ & $6.4\ ^{+0.2}_{-0.3}$ & $8.46\pm 0.01$ & 69    & $2.3^{a}$   \\

B0621-04 & 1.0391 & $8.3 \times 10^{-16}$ & $25     \ ^{+14}_{-16}  $ & $2.055\pm0.001$ & - & - & 32 & $0^{a}$   \\

B0809+74 & 1.2922 & $1.7 \times 10^{-16}$ & $-16   \ ^{+1}_{-16}   $ & $11.1\pm 0.1$ & $-13.2\ ^{+0.1}_{-0.7}$ & $11.12\pm 0.01$ & 9       & $4.5^{a}$   \\

B0818-13 & 1.2381 & $2.1 \times 10^{-15}$ & $-6.5   \ ^{+0.2}_{-0.7}$ & $4.7\pm 0.2$ & $-5.1\ ^{+0.1}_{-0.6}$ & $4.74\pm 0.01$ & 15.5 & $5.1^{a},\ 2^{b}$   \\

B1702-19 & 0.2990 & $4.1 \times 10^{-15}$ & $-80    \ ^{+6}_{-70}    $ & $11.0\pm 0.4$ & $-90\ ^{+40}_{-50}$ & $10.8\pm 0.2$ & 85 & $-4.1^{a},\ 4.1^{b}$   \\

B1717-29 & 0.6204 & $7.5 \times 10^{-16}$ & $-9.6   \ ^{+3}_{-0.6}   $ & $2.45\pm 0.02$ & $-10.9\ ^{+0.4}_{-0.7}$ & $2.461\pm 0.001$ & 28.9 & $4.6^{b}$   \\

B1844-04 & 0.5978 & $5.2 \times 10^{-14}$ & $80      \ ^{+70}_{-45}  $ & $12   \pm 1$ & - & - & 23 & $4.1^{a}$   \\

B2045-16 & 1.9616 & $1.1 \times 10^{-14}$ & $17      \ ^{+18}_{-2}    $ & $3.2\pm 0.1$ & $-26\ ^{+4}_{-2}$ & $3\pm 0.1$ & 36 & $1.1^{a},\ 1.1^{b}$   \\

B2303-30 & 1.5759 & $2.9 \times 10^{-15}$ & $15      \ ^{+3}_{-0.3}   $ & $2.1 \pm 0.1$ & $10.6\ ^{+0.8}_{-0.2}$ & $2.06\pm 0.02$ & 20.5 & $4.5^{a}$   \\

B2310-42 & 0.3494 & $1.1 \times 10^{-16}$ & $60      \ ^{+20}_{-10}$ & $2.1\pm 0.1$ & $13\ ^{+4}_{-6}$ & $2.1\pm 0.05$ & 56 & $6.8^{a}$   \\

B2319+60 & 2.2565 & $7.0 \times 10^{-15}$ & $70     \ ^{+60}_{-10}$ & $7.7\pm 0.4$ & $80\ ^{+30}_{-20}$ & $5\pm 3$ & 18 & $2.2^{a},\ 2.3^{b}$   \\[1ex]
\hline
\end{tabular}
\label{pulsars}
%\end{widetext}
\end{table*}

%%%%%%%%%%%%%%%%%%%%%%%%%%%%%%%%%%%%%%%%%%%%%%%%%
\section{Comparison with the observed velocities of the drifting subpulses}
\label{alt}
%%%%%%%%%%%%%%%%%%%%%%%%%%%%%%%%%%%%%%%%%%%%%%%%%

Equations (\ref{omegaDlow}) and (\ref{omegaDhigh}) give
the subpulse drift velocity within the SCLF model. In
order to check whether these expressions predict numbers
in agreement with observations,
we have used the data from the catalog of \cite{wes06} and \cite{wse07}.

These authors present the results of the observations of
187 pulsars in the northern hemisphere at wavelengths of
$21\, \mathrm{cm}$ and $92\, \mathrm{cm}$. Pulsars revealing
the drifting subpulses phenomenon are divided into three
classes, depending on the character of their
Two-Dimensional Fluctuation Spectrum (2DFS). Coherent
drifters  (marked as Coh) have narrow pronounced feature
in their 2DFS spectra, meaning that $P_3$ has a stable
value through the observations. Diffuse drifters of the classes Dif
and  Dif$^*$ have a broader feature in 2DFS spectra,
which for Dif pulsars is clearly separated from the alias
borders of the spectra ($P/P_3=0$ and $P/P_3=0.5$),
while for Dif$^*$ pulsars is not (see \cite{wes06} for
more details and examples).
For our purposes
we considered the coherent drifters from Tab.~2
of \cite{wes06} (corresponding to the observations at $21
\mathrm{cm}$), selecting only those with known
inclination angle $\chi$. The values of the inclination angles
were taken from \cite{r93} and, if absent there, from
\cite{lm88}. The resulting sample
is reported  in Table~\ref{pulsars}. When two values of
the periods $P_2$ and $P_3$ were given in \cite{wes06},
we chose the first one.

We assumed all pulsars to have the typical numbers for
compactness $\varepsilon = 0.4$, $\kappa=0.15$, and stellar radius
$R=10^6 \mathrm{cm}$. A clear picture of drifting subpulses is
observed when the line of sight of the observer grazes the
emission cone, so that it is reasonable to take $\xi = 0.9$. One
may also notice that the second term in (\ref{omegaDlow}) and
(\ref{omegaDhigh}), containing the dependence on $\phi$, is
smaller than the first term (it depends on a higher degree of the
small angle $\Theta_0$) and plays a role mostly for the pulsars
with large inclination angles. Hence, for the purposes of this
subsection we fixed\footnote{In any case we have verified
  that the
  results depend weakly on the
  parameters $\xi$ and $\phi$.} $\phi=\pi$.
Under these assumptions, the drift velocities (\ref{omegaDlow}) and (\ref{omegaDhigh}) for each individual pulsar depend only on the radial coordinate $\bar{r}$. So, by solving numerically the equation
\begin{equation}
\label{altitude}
\omega_{D\ \rm low/\rm high} = \frac{P_2}{P_3}
\end{equation}
for each pulsar of Table \ref{pulsars}, we can find
the altitude $\bar{r}_0 - 1$ of the plasma features that are responsible
of the subpulses.
When
solving the equation (\ref{altitude}) in the low altitude
approximation we took the first 30 terms of the expansion
(\ref{Philow}),
which reduces the error to less than one percent.

In Table \ref{pulsars} we have pulsars with both positive and
negative subpulse drift velocities, and no preferred direction of
the drift (sign of $P_2$) was found in \cite{wes06} and
\cite{wse07}. For comparison, we report in the table also the
values of the impact angle $\beta$, taken from \cite{r93} and
\cite{lm88}, which is the angle of the closest approach between
the magnetic axis and the line of sight. As we already mentioned,
the SCLF model predicts negative drift for the outer line of sight
(positive $\beta$) and positive drift for the inner line of sight
(negative $\beta$). However, from Table \ref{pulsars} we don't
see a correlation between the signs of $\beta$ and $P_2$.
According to the vacuum/partially screened gap model, the
discrepancy between the predicted and the observed direction of
the drift is usually explained in terms of aliasing
\citep{ggks04}.

One may assume that both the aliasing
  and the orientation of the 
line of sight may serve as an explanation of the visible drift direction,  
depending on the individual pulsar properties. However, some 
individual pulsars have non-trivial subpulse behavior, which can 
be a challenge for all existing models. There are pulsars showing
different sign of the drift velocity in different modes, or in the
same mode, like J0815+09. Six pulsars, having opposite 
drift senses in different components are present in the catalog of 
\cite{wes06}. The pulsar B2045-16, reported in  Table \ref{pulsars},
has opposite senses of the subpulse
drift for the observations at $21\, \mathrm{cm}$ and $92\,
\mathrm{cm}$. At least six more pulsars among those reported by
\cite{wes06} and \cite{wse07} show a similar  phenomenology.
However, all of them are diffuse drifters (class $\mathrm{Dif}^*$)
at $92 \, \mathrm{cm}$ (including B2045-16), suggesting
that aliasing is very likely to occur for them.

For the purpose of our analysis we have changed the sign of all considered 
subpulse drift velocities from Table \ref{pulsars} to negative. As a justification
for this we may point on the uncertainty in the determination of the observing geometry
(in different sources one may frequently find different 
estimations for the impact angle of the same pulsars).
At the same time we leave space for the existence of other yet unknown reason, 
responsible for the visible direction of the drift.
In this context the question which we rise 
is the following: are the typical values for
the subpulse drift velocities of different pulsars in general compatible with the 
predictions of the SCLF model?

\begin{figure}
\includegraphics[width=0.45\textwidth]{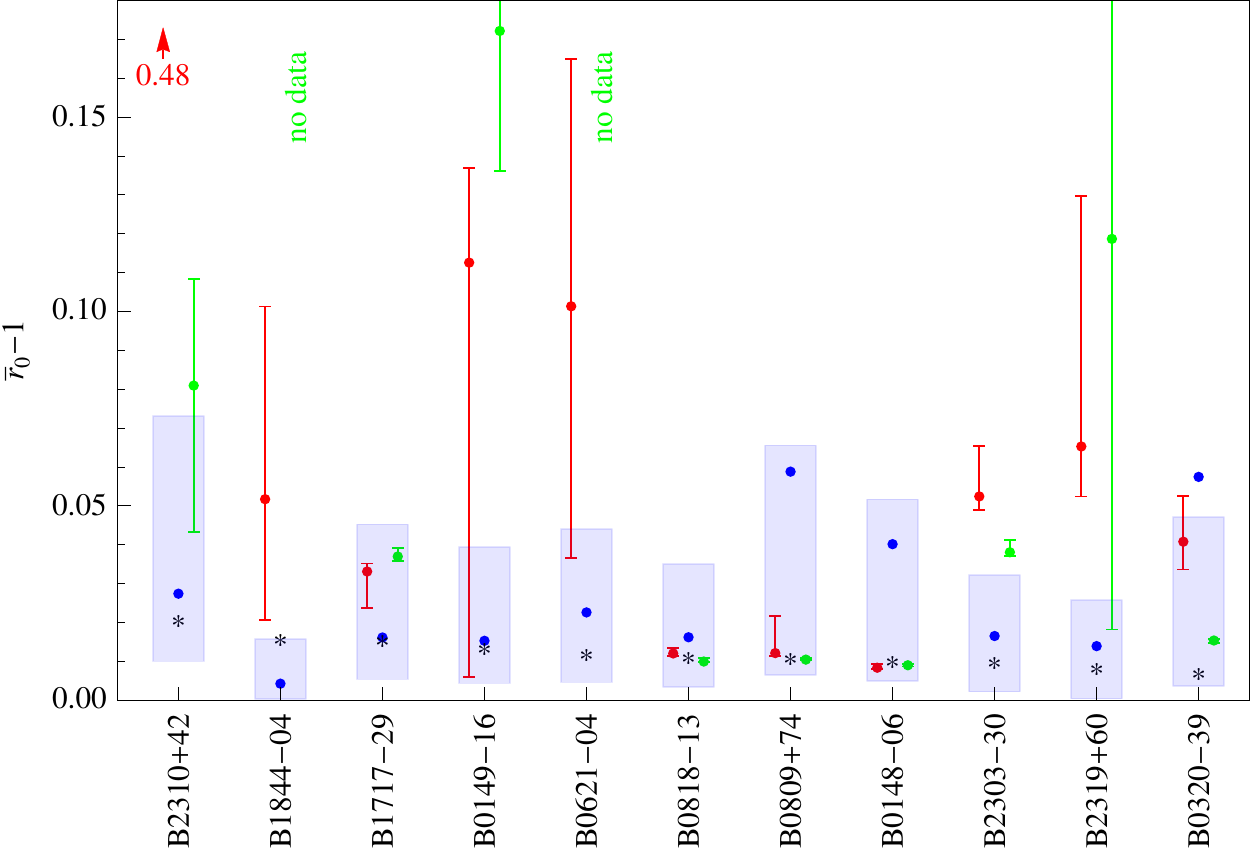}
\caption{Values of the altitude above the surface of the
  star $\bar{r}_0-1$ obtained by solving the equation
  (\ref{altitude}) for the pulsars of Table
  \ref{pulsars}. Red points correspond to the observing
  wavelength at 21 cm, green points correspond to the
  observing wavelength at 92 cm. The obtained altitudes
  are compared to the PFF from \citet{ha01} (blue
  shadowed regions) and \citet{hm01,hm02} (blue
  points). Black stars show the angular radius of the
  polar cap $\Theta_0$. Special cases are indicated as described in the text.
The figure does not
    report the two sources B1702-19 and B2045-16. The
    first one
does not admit a solution of Eq.~(\ref{altitude})  for any
radial coordinate, while the second one has
opposite sense of the subpulse drift for the observations
at $21\, \mathrm{cm}$ and $92\, \mathrm{cm}$.} \label{fig1}
\end{figure}

The results of our analysis are schematically represented in
Fig.~\ref{fig1}. The red points represent the values of the
altitude above the surface of the star in units of
stellar radii ($\bar{r}_0 - 1$) obtained from
Eq.~(\ref{altitude}), with the error bars calculated from
the errors of $P_2$ and $P_3$ given in Table 2 of
\cite{wes06} (at $21 \,\mathrm{cm}$). The green points
represent the same quantity obtained from the
observations at $92 \,\mathrm{cm}$ (\cite{wse07}). The
pulsars are arranged in ascending order of the
period. Black stars show the angular radii of the polar
caps at the surface of the star, $\Theta_0$, for
comparison.
Pulsars
which are absent in the $92 \, \mathrm{cm}$ catalog are
marked with ``no data''.  Arrow indicates the value of
the altitude too large to fit in the plot.

For the sake of comparison, we have also computed the
altitudes as given by
the heights of the pair formation front (PFF) following
\citet{ha01,hm01,hm02}. The expressions for the PFF location
in the framework of the SCLF model are given in
the Appendix.
We recall that the main processes determining
the height of the PFF in pulsar magnetosphere are
curvature radiation and inverse Compton scattering
(resonant as well as non-resonant). In Fig.~\ref{fig1}
the shaded blue regions indicate the PFF altitude from
\cite{ha01} for the curvature radius of the surface
magnetic field lines starting from the values $\sim R$
(lower boundary of the shaded zones) 
up to the values of
the standard dipolar magnetic field (upper boundary of
the shaded zones). The blue points show the PFF altitudes
from \cite{hm01,hm02}. The temperature of the star is taken to be $T=2\times 10^6 \mathrm{K}$.

Even though our approach fails for some of the considered
pulsars, we find it very promising that for most of them the
altitude corresponding to the observed value of the
subpulse drift velocity lies within the range or slightly
above the PFF altitudes. It is a well-known fact that radio
emission in pulsar magnetosphere originates at heights of
the order of tens of stellar radii
\citep{kx97,k01,kg03}. However, it was assumed in
\cite{rs75} and subsequent works on drifting subpulses
that the geometrical pattern and drift velocity of the
subpulses are determined by the distribution of sparks
within the gap, \ie very close to the surface of the
star. In this respect we think that the heights that we
have obtained, so close to the PFF values, are not occasional and may
reflect the fact that, whatever is the mechanism of
subpulse generation in the SCLF model, they are likely to
form in the vicinity of the pair formation front.

Our analysis was not meant to consider systematically all the
sources in the catalog of \citet{wes06}, but rather it was aimed at
demonstrating the potential ability of the model in accounting for
the observations. While we are confident that the main conclusion
of this subsection may be applicable also to other pulsars, a
systematic study of all the sources may
become necessary.

%%%%%%%%%%%%%%%%%%%%%%%%%%%%%%%%%%%%%%%%%%%%%%%%%
\section{Discussion of specific sources}
\label{individualpulsars}
%%%%%%%%%%%%%%%%%%%%%%%%%%%%%%%%%%%%%%%%%%%%%%%%%

The expressions for the subpulse drift velocity
(\ref{omegaDlow}) and (\ref{omegaDhigh}) derived in the
framework of the SCLF model naturally contain a
dependence on $\xi$ and
$\phi$, and they predict different velocities for
different regions of the polar cap, in contrast to the
estimations of the vacuum gap model.
In this subsection we attempt to exploit these
  additional degrees of freedom
  to explain the
  variability of the subpulse velocities along the pulse
  longitude in case of two specific  pulsars, \ie
PSR B0826-34 and PSR B0818-41. Although not included in
the catalog of \citet{wes06},
both of them
have been  repeatedly investigated
at several observing frequencies, and, since they have wide
profiles allowing to track several subpulse drift bands
at a time, they can be regarded as ideal test cases.

%%%%%%%%%%%%%%%%%%%%%%%%%%%%%%%%%%%%%%%%%%%%%%%%%
\subsection{PSR B0826-34}
\label{B0826-34}
%%%%%%%%%%%%%%%%%%%%%%%%%%%%%%%%%%%%%%%%%%%%%%%%%

\subsubsection{Basic parameters}
The pulsar B0826-34, with spin $P=1.8489 ~\mathrm{s}$ 
and $\dot{P}=1.0\times 10^{-15}$, has an
unusually wide profile, extending through the whole pulse period.
The pulsar emits in its strong mode for $30\%$ of the time. For
the rest of the time, the pulsar stays in the weak mode, with an
average intensity of emission which is $\sim 2\%$ of the emission
of the strong mode \citep{eetal05,s11,eamn12}. Because of its
weakness, the very existence of the weak mode was confirmed only
very recently and for a long time it was thought to be a null
pulsar \citep{detal79,betal85,bgg08}.

The average pulse profile of B0826-34 consists of the
main pulse (MP) and the interpulse (IP), separated by
regions of weaker emission. The intensity of the MP is
much larger than the intensity of the IP at the
frequencies $318\, \mathrm{MHz}$ and $606\, \mathrm{MHz}$, while at
frequencies larger than $\sim 1 \mathrm{GHz}$ the IP
starts to dominate. The MP itself has a double peaked
structure with a separation between the peaks
decreasing at higher frequencies, following the common
trend described by the radius-to-frequency mapping model
of \citet{kg03}. A detailed description of the average
profile evolution with frequency may be found in
\cite{ggks04}, \cite{bgg08}.

Additional relevant physical parameters are those related
to the observing geometry of PSR B0826-34, \ie the
values of the inclination angle $\chi$ and of the impact
angle $\beta$. Usually the values of these angles are
determined by fitting the polarization profile of the
pulsar. According to the ``rotating vector model''  of \citet{rc69}, the polarization angle of the pulsar radio emission is equal to
\begin{equation}
\label{PA}
\phi_{PA} = \tan^{-1}\frac{\sin\chi\sin l}{\sin(\chi+\beta)\cos\chi-\cos(\chi+\beta)\sin\chi\cos l}\,,
\end{equation}
where $l$ is the pulse longitude, related to the
azimuthal coordinate $\phi$ by means of standard theorems of
spherical geometry
\citep{ggks04}\footnote{In the corresponding formula of
  \cite{ggks04} the azimuthal coordinate with respect to
  the magnetic axis is denoted by $\sigma$, while $\phi$
  is used to denote the pulse longitude, which may
  introduce a confusion when making comparison with our results.
  However, we preferred to keep the notation $\phi$ for the azimuthal angular coordinate as in \cite{mt92}. } as
\begin{equation}
\label{l}
\sin l = \frac{\sin\phi \sin[\xi \Theta]}{\sin(\chi + \beta)}\ .
\end{equation}
However, in many cases this method does not give a unique
value for the inclination and for the impact angles, but
rather a wide range of possible
combinations~\citep{mh93}. For example, early estimations
of \cite{betal85} for PSR B0826-34 based on the polarization
measurements suggested a large range for $\chi$ and
$\beta$ with the best fit of $\chi =
53^\circ \pm 2^\circ$ and $\beta + \chi = 75^\circ \pm
3^\circ$, a fact which does not agree with
the large width of the profile.
In \cite{ggks04} these angles were estimated
from the polarization information\footnote{
We recall that the measurements of linear polarization are
sometimes contaminated by the orthogonal polarization mode
switching. Single pulse polarization observations may
be necessary for the reconstruction of the mode-corrected
polarization angle swing \citep{gl95}.
}
 together with the
frequency evolution of the profile and were found to lie in the
range\footnote{ These ranges include the values reported by
\cite{lm88} $\chi=2.1^\circ $ and $\beta =1.2^\circ $ and by
\cite{r93} $\chi = 3^\circ$ and $\beta = 1.1^\circ $. On the other
hand, \cite{eetal05} report the value of $\chi = 0.5^\circ $. }
$1.5^\circ \leq\chi\leq 5.0^\circ $ and $0.6^\circ  \leq\beta\leq
2.0^\circ $.

\subsubsection{Subpulse phenomenology}
The drifting subpulses are seen almost along the whole
range of longitudes, showing from 5-6 up to 9 visible
tracks at a time. The character of the observed subpulse
drift is irregular, the apparent velocity reveals an
oscillatory behavior, changing sign with a periodicity of
the order of tens to several hundred periods of the
pulsar. The values of the apparent subpulse drift
velocities were measured in a number of papers for
different observing frequencies
\citep{ggks04,betal85,eetal05,lt12} and are reported in
Table~\ref{velocity}. In all cases the declared velocity
range is not symmetric with respect to zero, having a
positive average velocity
(see Figs. 5 and 6 of \cite{eetal05} for an example of a
positive average drift).

\begin{table*}
%\begin{widetext}
\centering
\caption{Subpulse drift velocity ranges of PSR B0826-34 measured at different observing frequencies.}
\begin{tabular}{c c c c}
\hline\hline
Observing frequency (MHz) & Measured drift velocity ($^\circ/P$) & Reference & Average drift velocity ($^\circ/P$)  \\ [0.5ex] % inserts table %heading
\hline
318 & $-0.8 \div 1.9$ & \cite{ggks04} & $0.55$   \\
645 & $-1.5 \div 2.1$ & \cite{betal85} & $0.3$ \\
1374 & $-3.2 \div 3.6$ & \cite{eetal05} & $0.2$ \\
1374 & $-1 \div 1.5$ & \cite{lt12} & $0.25$  \\[1ex]
\hline
\end{tabular}
\label{velocity}
%\end{widetext}
\end{table*}

Another interesting property of PSR B0826-34 is the longitude
dependence of the period $P_2$. \cite{eetal05} reports the values
for $P_2$ between $26.8^\circ$ and $28^\circ$ in the MP region
(average $27.5^\circ$) and between $19^\circ$ and $23.5^\circ$ in
the IP region (average $22.2^\circ$). In order to explain this
behaviour the authors proposed a model of spark carousel
consisting of two rings of 13 sparks each, with the separation
between the sparks in the outer ring (responsible for the MP)
$27.5/22.2\approx 1.2$ larger than in the inner ring (responsible
for the IP). In \cite{ggks04} $P_2$ was found to vary between
$21.5^\circ$ and $27^\circ$ with the mean value of $24.9^\circ$.
As a possible explanation of the observed phenomena the authors
proposed a scheme where the ring of sparks is centered not around
the dipolar axis of the pulsar magnetic field, but around the
``local magnetic pole'', shifted with respect to the global dipole
one. This agrees well with the models suggesting that the pulsar
magnetic field has a multipole structure near the  surface of the
neutron star \citep{gs00}, which arises due to the dynamo
mechanism in the newborn stars \citep{ug04} or, more probably, due
to the Hall drift \citep{grg03,ggm13}.

\subsubsection{Analysis of the subpulse drift}

We start our analysis of the subpulse drift by estimating the
altitude above the surface of the star, which would correspond to
the average subpulse drift velocities of PSR B0826-34 [\cf
Sec.~\ref{alt}]. Positive average observed drift velocities contradict 
our equation (\ref{omegaDlow}), which predicts negative values for the 
drift velocity everywhere across the polar cap of a nearly aligned pulsar. 
Since in the case of PSR B0826-34 the observer's 
line of sight lies most probably entirely
in the polar cap region, one cannot explain the positive
observed drift velocity with  the negative impact angle $\beta$. 
However, one can suppose that, rather than
drifting in the positive direction with some small velocity $\omega$, the 
subpulses actually drift in the negative direction, in
such a way that, at every period, 
each successive subpulse appears close to the place of the preceding one.
For example, taking the average value of the subpulse drift velocity from \cite{ggks04} as 
$0.55^\circ / P$ and using the average value of the
subpulse separation $P_2$ estimated there as
$24.9^\circ\pm0.8^\circ$, one can see that similar observed
picture would be obtained  
if the subpulses were drifting with the negative
velocity
$(0.55^\circ - 24.9^\circ)/P = -24.35^\circ / P$,  
provided the period $P_3$ is close to the pulsar period $P$. 
The period $P_3$ of PSR B0826-34 is not yet
reported in the literature, possibly because the subpulse
tracks are irregular. Hence, the lack of observational indications
legitimate us to assign
any reasonable value to $P_3$. For instance, 
in the work of \cite{ggks04} good fits of the observational data 
are obtained when the values of $P_3$ are 
equal to $1.00 P$, $0.5 P$ and $0.33 P$. For the purposes of our analysis, 
in this subsection we will assume that $P_3\sim P$ and
that all the observed 
drift velocities are in fact shifted by $-24.9^\circ / P$, so that the altitude 
of the plasma features responsible for the subpulses in case of PSR B0826-34 
should correspond to the average plasma drift velocity $- 24.35 ^\circ / P$. Assuming
that the pulsar is nearly aligned, the best result is obtained for 
$\bar{r}_0 - 1=0.19$. This altitude depends weakly on the
chosen angle $\chi$, provided the latter is close to
$0^\circ$. This value is somewhat higher than the predicted values
for the PFF, which for our reference temperature 
$T=2\times 10^6 \mathrm{K}$ are $0.005<h^{\rm
HA}<0.059$ (depending on the value of $f_{\rho}$) 
and $h^{\rm HM}=0.062$ (see the Appendix for the definition of $h^{\rm
HA}$ and $h^{\rm HM}$). 
However, one may notice that 
the altitude of the PFF is quite sensitive to the temperature of the star
(in case when it is controlled by the inverse Compton scattering), and, for instance, for the 
temperature $T=0.5\times 10^6 \mathrm{K}$ the corresponding
values for $h^{\rm HA}$ lie in the range $0.021-0.235$, with the same
$h^{\rm HM}$.

\begin{figure}
\begin{center}
\includegraphics[width=0.35\textwidth]{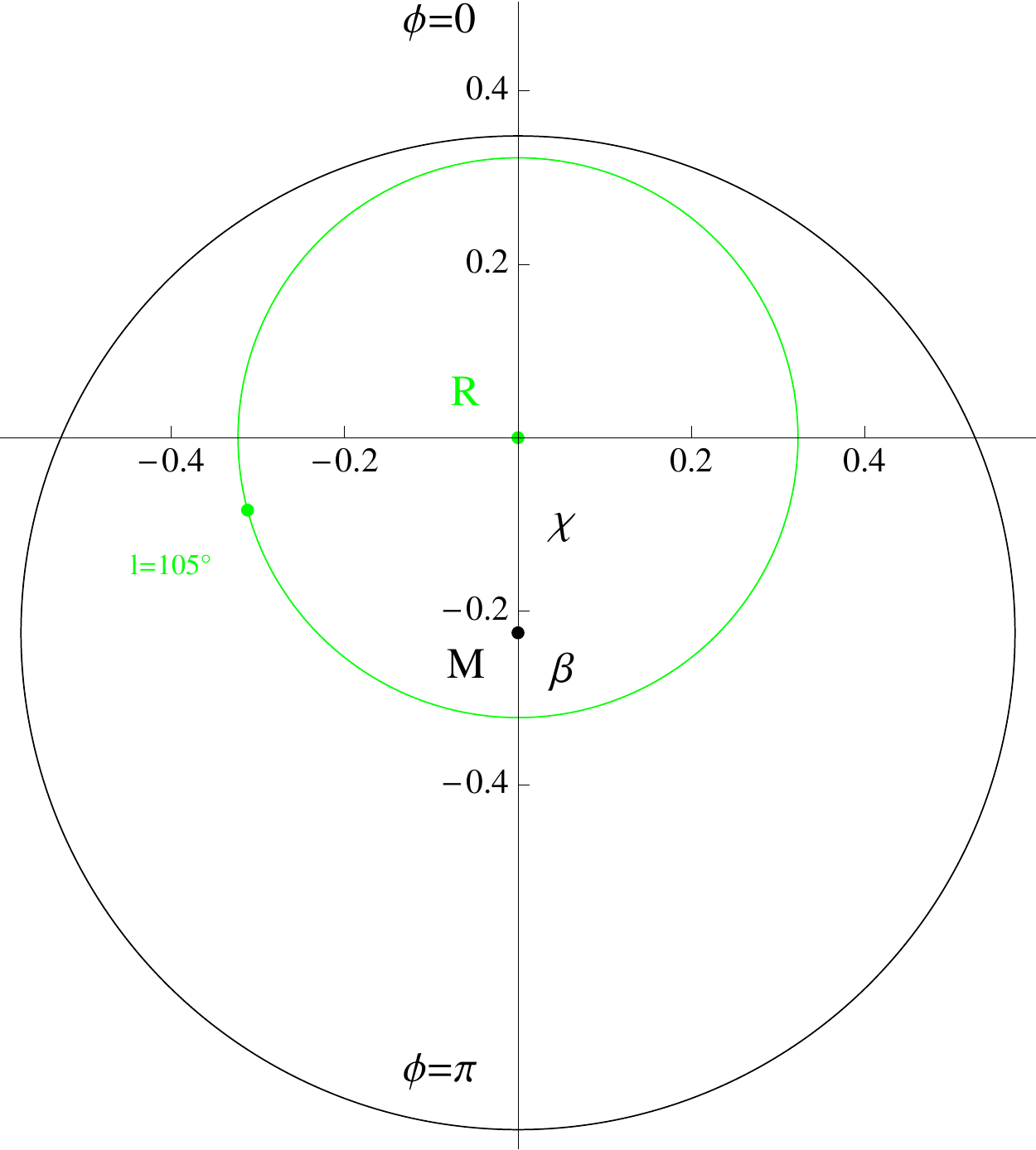}
\includegraphics[width=0.49\textwidth]{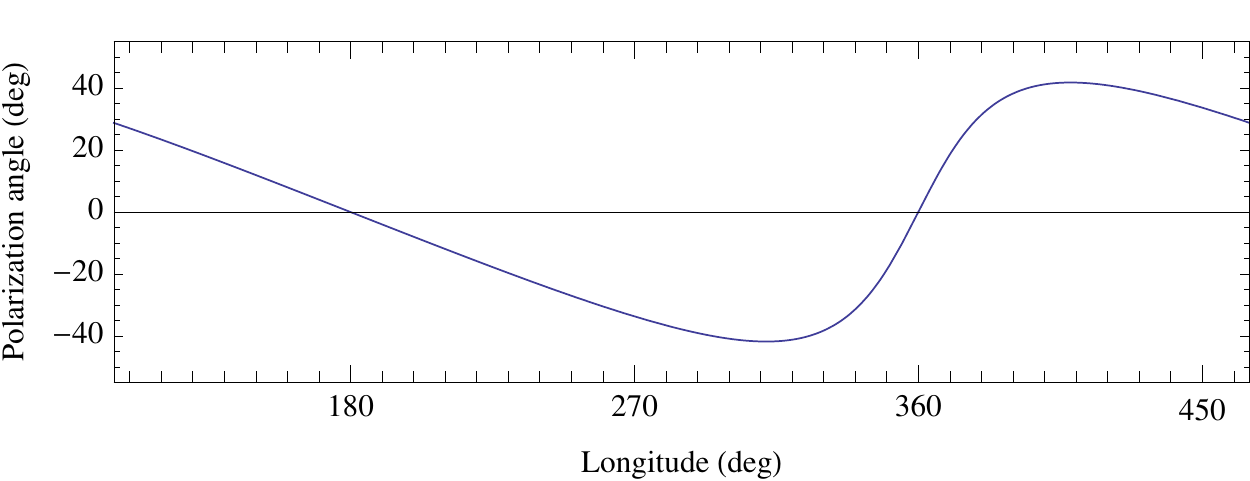}
\end{center}
\caption{Upper panel: observing
  geometry of PSR B0826-34 (the coordinate axes show the
  values of the angular coordinate $\theta$ in
  degrees). The black circle indicates the polar cap,
  with the center in the magnetic pole (black point "M"),
  while the green circle indicates the trajectory of the
  line of sight of the observer, with the center on the axis 
  of rotation (green point "R"). "$l=105^\circ$" 
  (rather than $l=0^\circ$) is used 
  as origin of the pulse longitude in order to bring the 
  plots in the visual correspondence 
  with the ones from the literature. 
  Lower panel: Polarization angle as a function
  of longitude in our model. The plot is shifted in
  longitude for easier comparison with previous
  results from the literature (see discussion in the
  text).
} \label{PC}
\end{figure}

Fig.~\ref{PC} is devoted to the illustration of the geometry of
PSR B0826-34. In the upper panel we show a sketch of the polar
cap, whose boundary is represented with a black circle. At the
altitude $\bar{r}_0-1=0.19$, the polar cap has an angular radius
$\Theta\sim 0.57^\circ$ [\cf Eq.~(\ref{theta0})]. It should be
stressed that different observations of the angular size of the
polar cap do not provide consistent conclusions. According to
\cite{lm88}, for instance, $\Theta\sim 13^\circ P^{-1/3} /2$,
which, for PSR B0826-34, gives $\Theta\approx 5.3^\circ$, \ie an
order of magnitude larger compared to what we have found.
On the contrary, the angular size of the polar cap deduced from
X-ray observations is much smaller and close to the values that we
have obtained through Eq.~(\ref{theta0}).

Another relevant quantity is
the trajectory of the observer's line of sight, which  is given by~\citep{mt77}

\begin{widetext}
\begin{equation}
\xi = \frac{1}{\Theta(\bar{r})}\sin^{-1}\left[\frac{\cos(\chi+\beta)\cos\chi - \sin\chi\cos\phi \sqrt{\sin^2(\chi + \beta)-\sin^2\phi\sin^2\chi}}{1-\sin^2\chi\sin^2\phi}\right]\,,
\end{equation}
\end{widetext}
with $0\leq\phi<2\pi$, and is
represented with a green line in the upper panel of
Fig.~\ref{PC}.
The range of the coordinate $\xi$ along
  the line of sight approximately coincides with that
  estimated in \cite{ggks04} as $0.25 \div 0.85$.
In agreement with the tiny size of the polar cap,
we have chosen the inclination angle and the impact angle as
$\chi=0.225^\circ$ and $\beta=0.098^\circ$.
The resulting ratio
is consistent with the available observational data,
which uses the relation $\chi/\beta
\approx \sin\chi/\sin\beta=(d\psi/d\phi)_{max}$, where
$(d\psi/d\phi)_{max}$ is the value of the steepest
gradient of the polarization angle curve, estimated to be
$2.0^\circ  \pm 0.5^\circ  /^\circ$ in \cite{ggks04} and
$1.7^\circ  /^\circ$ in \cite{lm88}\footnote{This makes
our picture for the observed geometry a bit different
from Fig.~10 of \cite{eetal05}, where the angle $\beta$
is apparently larger than $\chi$.}.
The lower panel of Fig.~\ref{PC} reports the
  polarization profile computed through Eq.~(\ref{PA}).
  It should be compared with Fig.~6d of \cite{betal85}
  and with the upper panel of Fig.~1 of
  \cite{ggks04}\footnote{\label{fn}Note that the plots
  of the polarization angle and of apparent drift velocity
  in this subsection are shifted in longitude in order to
  make them easier comparable with the corresponding
  plots in the literature, associating zero of the pulse
  longitude with the bridge region before the IP. 
  In our analysis the zero of the azimuthal angle and of the
  pulse longitude is instead associated with the peak of the
  MP. The shift is taken to be $105^\circ$ to match the
  distance between the second zero of the position angle
  curve and the end of the pulse from the top panel of
  Fig.1 of \cite{ggks04}.}.
  
\begin{figure}
\begin{center}
\includegraphics[width=0.45\textwidth]{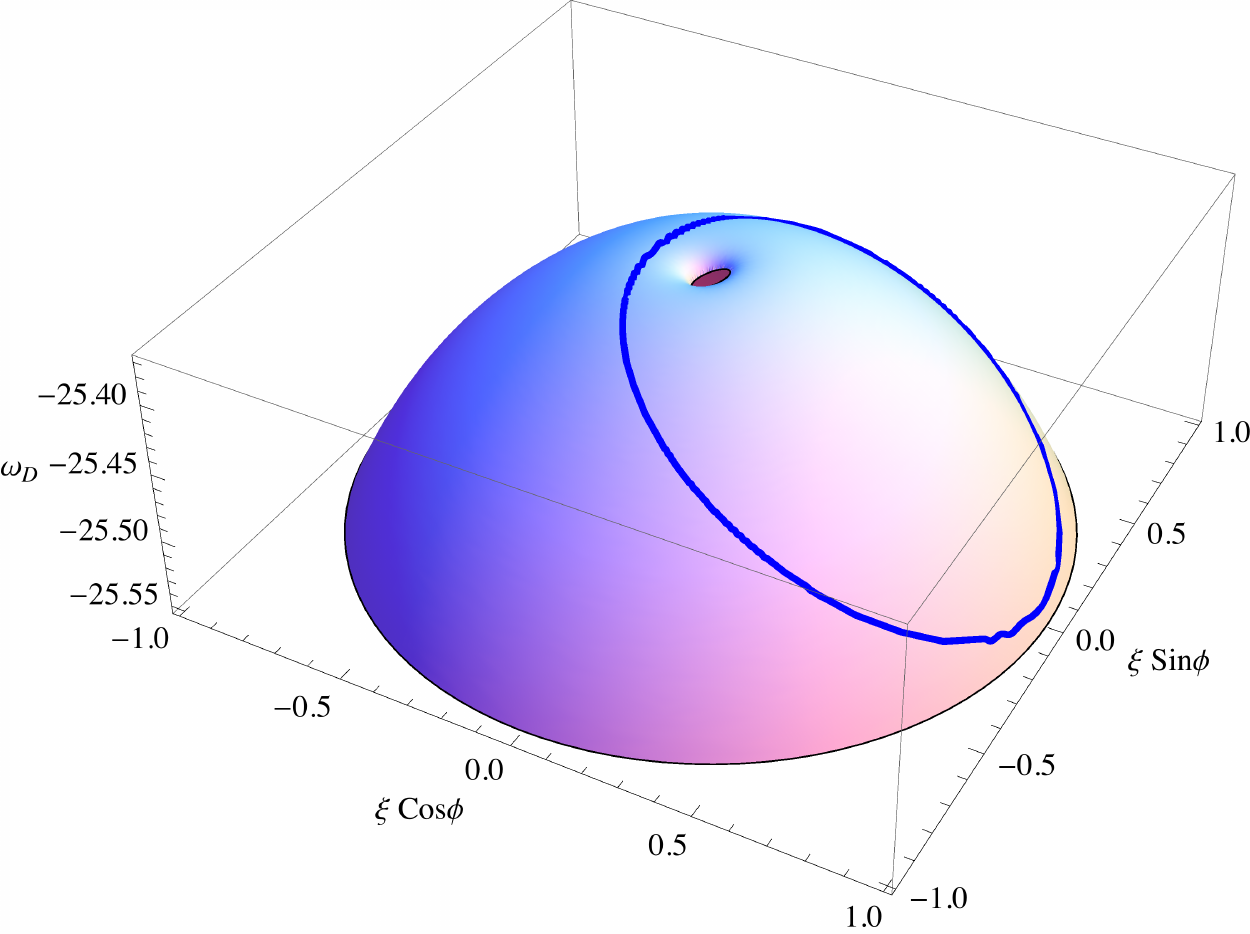}
\includegraphics[width=0.49\textwidth]{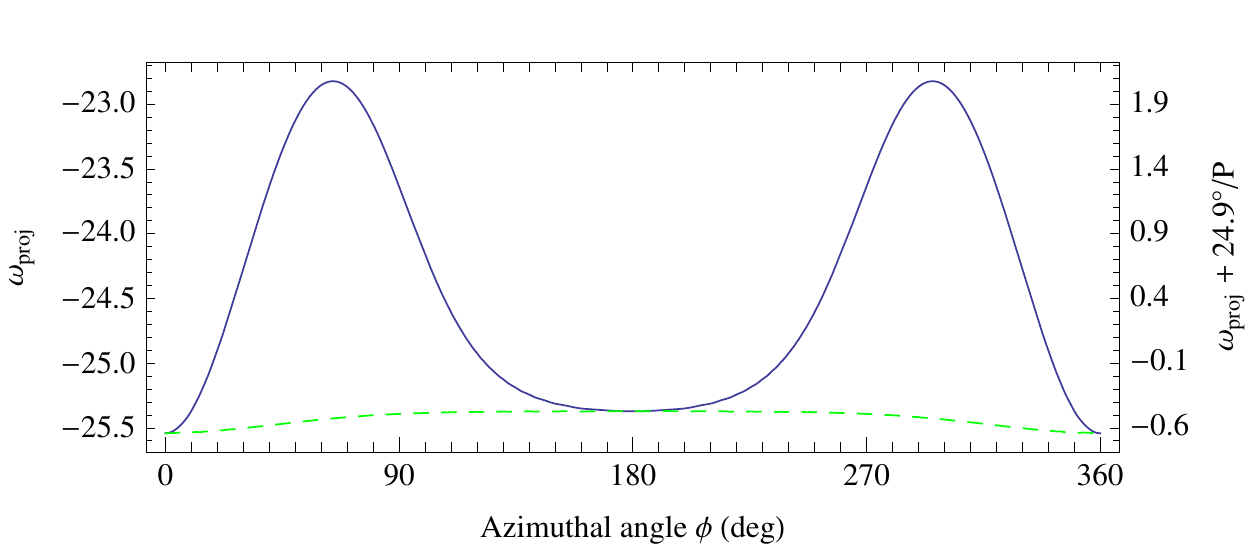}
\end{center}
\caption{Upper panel: 3D visualization of the plasma drift velocity
across the polar cap ($0<\xi<1$, $0<\phi<2\pi$) of a pulsar with the 
parameters of PSR B0826-34 (the radial coordinate is taken to be 
$\bar{r}=1.19$). The solid blue line shows the observer's line of sight 
according to the model presented in Fig.~\ref{PC}.
  Lower panel: The dashed line represents the plasma drift velocity 
  $\omega_{D\  low}$ along the line of sight of the
  observer. The blue solid
  line shows the projection of the plasma drift velocity on the trajectory of the 
  line of sight (actually measured velocity). The values reported on the 
  right vertical axis represent the drift velocity 
  shifted by the constant value $24.9^\circ / P$, as it appears to the 
  observer.
} \label{Vel3D}
\end{figure}

The upper panel of Fig.~\ref{Vel3D} shows the 
3D plot of the plasma drift velocity 
(\ref{omegaDlow}) in the polar cap region ($0<\xi<1$, 
$0<\phi<2\pi$) for the set of 
parameters of PSR B0826-34 and for the radial coordinate $\bar{r}=1.19$\ \ 
 \footnote{The first 200 terms of the infinite series of the 
expression (\ref{omegaDlow}) are plotted.}.
The expression (\ref{omegaDlow}) diverges for the values of $\xi$ 
close to zero and we cut this region from the plot. One may notice
that, in spite of the complexity of  the expression (\ref{omegaDlow}), 
the resulting angular dependence of the 
plasma drift velocity is quite smooth. The blue solid 
line shows the line of sight of the observer, 
which 
corresponds to the green circle in Fig.~\ref{PC}.
In our model we associate the region around $\phi=0$ to the MP 
of PSR B0826-34, while the region around $\phi=\pi$ is associated to the IP.
The lower panel of Fig.~\ref{Vel3D}  shows the drift velocities along the line of sight
of the observer. The green dashed line shows the actual values of the 
plasma drift velocity $\omega_{D\ low}$ (left vertical
axis) at the points crossed by the line of sight
as a function of the azimuthal angle $\phi$.
The blue solid line shows
the drift velocity which the observer will actually measure, i.e. the
drift velocity (\ref{omegaDlow}) projected on the trajectory of
the line of sight across the polar cap
\begin{equation}
\label{omegaproj}
\vec{\omega}_{proj}=\frac{\vec{\omega}}{\sqrt{1+\left(\frac{d\xi}{d\phi}\right)^2}}\,.
\end{equation}
This is again plotted 
as a function of the azimuthal angle $\phi$, 
and  the green and the
blue curve coincide at $\phi=0$ and $\phi=\pi$,
as they should.
The values indicated on the right vertical axis
represent the projected velocity $\vec{\omega}_{proj}$ shifted by the
constant value $24.9^\circ / P$ and give the values of the drift velocity, 
apparent to the observer.
Interestingly, one can see that these values cover the observed range of drift velocities 
reported by \cite{ggks04}, namely $(0.8 \div 1.9) ^\circ /
P$. This suggests
that the diversity in the measured velocity of PSR B0826-34 
may be explained with the intrinsic angular dependence of the plasma drift velocity across 
the pulsar polar cap.

Following a similar argument, 
we can try to explain 
the observed longitude dependence of the subpulse separation $P_2$ of 
PSR B0826-34.
The subpulse drift velocity is usually defined as 
$\omega = P_2/P_3$. However, one may alternatively assume
that the observed subpulse separation 
$P_2$ is, in fact, determined by the velocity $\omega$, which, 
in turn, is defined by the physical condition of the 
plasma at any given point of the polar cap.
Intuitively, if one imagines the carousel of sparks (or any other
feature responsible for the subpulse phenomena) moving around the
polar cap with the longitude dependent velocity, it seems
plausible that in the regions with lower velocity the subpulse
tracks will tend to look closer, while in the regions with larger
velocity they will look farther away from each other. From the 
observations of other pulsars we know
that the values of $P_3$, which, we recall, also enters the expression of the
apparent drift velocity, are essentially the same at different
observing frequencies for a given pulsar
\citep{wes06,wse07,bgg09}, suggesting that this subpulse parameter
expresses some persistent property.
We therefore argue that the longitude
dependence of $P_2$ is explained in terms of the longitude
dependence of the drift velocity of the features that
generate the subpulse,
while $P_3$ could represent a specific characteristic
property of the individual pulsar.

\begin{figure}
\begin{center}
\includegraphics[width=0.49\textwidth]{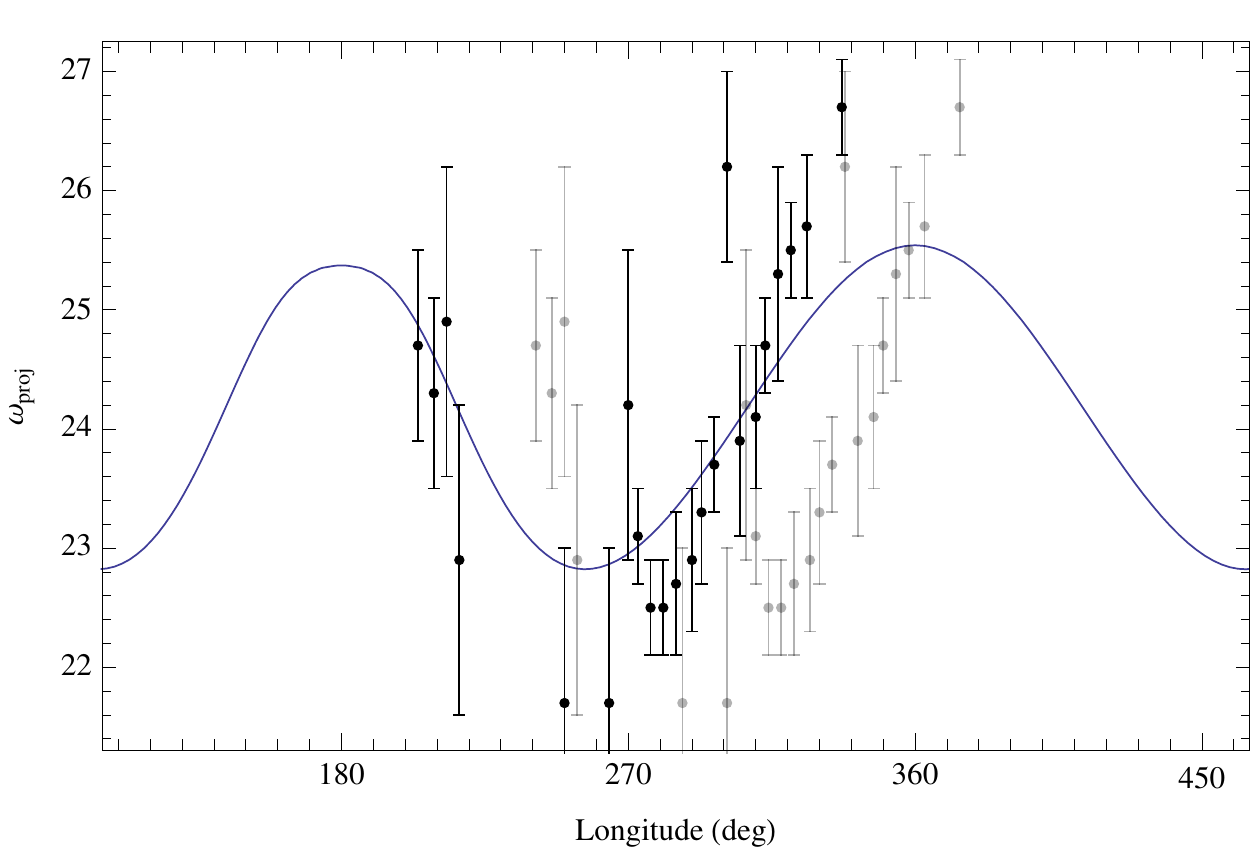}
\end{center}
\caption{Comparison of the variation of the plasma drift
  velocity along the pulse longitude (solid blue curve)
  with the corresponding variation of the period $P_2$ of
  PSR B0826-34, measured in \citet{ggks04}. The values of $P_2$ from the upper panel of the Fig.8 of \citet{ggks04} are shifted on $37^\circ$ (black points), to bring them in better visual correspondence with the velocity curve (the value of shift is not a result of fit, but estimated by eye).} \label{fit}
\end{figure}

In order to support this point of view, we have tried to make 
the analysis of the longitude
dependence of $P_2$ reported in \cite{ggks04} and our results
are reported in Fig.~\ref{fit}. The solid blue curve shows the absolute
value of the projected drift velocity (\ref{omegaproj}),
as a function of the pulse longitude $l$ \footnote{Note, that, apart for
the $105^\circ$ shift in the longitude, the shape
of the velocity in Fig.~\ref{PC}  is different from that
in Fig.~\ref{Vel3D},  since the pulse longitude 
is in general different from the azimuthal angle $\phi$.}. The grey points
represent the observed values of $P_2$ given in the top
panel of Fig.~8 of \cite{ggks04} along the pulse profile,
shifted by $105^\circ$ to bring them in correspondence with
our longitude scale. Note that, due to the adopted assumption $P_3\sim P$, 
these values of the pulse separation should reproduce the values of the
drift velocity along the pulse. The black points
are the same as the grey ones, but shifted by
$37^\circ$ in longitude. Although the value of this shift is
chosen ``ad hoc''  and is not the result of a fitting procedure,
there is a positive uncertainty in our choice of the origin of the
longitude (see the footnote \ref{fn}), as well as a possible
longitude shift between the $318 \mathrm{MHz}$ data of
\cite{ggks04} and the $606 \mathrm{MHz}$ data of \cite{betal85},
containing the polarization data that we have chosen as a reference here. 
In spite of the uncertainties, the correspondence that we have found between the
data points and the analytical curve is very promising and needs
further investigation and comparison with more data.

% -  (here the plus sign in front of the square root corresponds to the outer LOS geometry)

Finally, as already mentioned before, the measured subpulse drift
velocity of PSR B0826-34 reveals an irregular behavior on the
timescales of tens to hundreds of period, for which a firm
explanation is still lacking. Indeed, our basic model, 
described in Figs.~\ref{PC} and~\ref{Vel3D}, predicts a
certain value of the drift velocity for a certain value of the 
pulse longitude, while the observed sequences of pulses taken
from 
\cite{eetal05,ggks04,lt12} suggest that the velocity 
at a given pulse longitude changes with time.
\cite{ggks04} explained these
variations within the partially screened gap model by invoking
small fluctuations of the polar cap temperature around the mean
value\footnote{The key assumption of this interpretation is that
the visible subpulse velocity is in fact the aliased value of the
true (higher) one, so that the periodic change of the apparent
drift direction corresponds to the slowing down and speeding up of
the intrinsic drift rate with respect to its average value. }
\citep{gmg03}. On the contrary, \cite{lt12} argued that the
potential drop in the polar cap region may be determined not only
by the local physical conditions, but by the global structure of
the magnetosphere \citep{lst12,kkhc12,t10a,t10b}. As a result, the
long term changes of the observed drift rate may be related to the
evolution of the magnetospheric current density distribution, for
example, due to switches between meta-stable magnetosphere
configurations \citep{t10a}.

The alternative explanation that we may propose is that the observed
variations are related to stellar oscillations. In our
preceding research \citep{maz10,zma12} we have studied the
influence of the non-radial stellar oscillations on the scalar
potential of the polar cap region of the magnetosphere. The
oscillation velocity at the stellar surface modulates the linear
velocity of the pulsar rotation, introducing a new term in the
charge density, in the scalar potential and in the accelerating
electric field above the surface of the star. We also shown that
oscillations may increase the electromagnetic energy losses of the
pulsar, causing its migration above the death-line in the
$P-\dot{P}$ diagram. Taking into account that PSR B0826-34 is
located relatively close to the death-line ($\tau=3\times10^7
\mathrm{yr}$, $B_d=1.4\times 10^{12} \mathrm{G}$) and that it is
very intermittent, staying in the ON state for only $30\%$ of the
time, it is likely that this pulsar is visible only when it
oscillates, which may also determine the character of variation of
the observed subpulse velocity.\footnote{ The observed periodicity
of hundreds of seconds may be reached  by core g-modes of the
neutron star \citep{mhh88}.}

\

%%%%%%%%%%%%%%%%%%%%%%%%%%%%%%%%%%%%%%%%%%%%%%%%%
\subsection{PSR B0818-41}
\label{B0818-41}
%%%%%%%%%%%%%%%%%%%%%%%%%%%%%%%%%%%%%%%%%%%%%%%%%

\subsubsection{Basic parameters}

The pulsar B0818-41, with the main parameters $P=0.545\,\mathrm{s}$,
$\tau=4.57\times 10^8 \mathrm{yr}$, $B_d=1.03\times 10^{11}
\mathrm{G}$, was discovered during the second pulsar survey
\citep{metal78,hetal04}. The width of the average pulse profile of
PSR B0818-41 is close to $180^\circ$ with a pronounced subpulse
drift along a wide range of longitudes. The typical subpulse drift
pattern (see Fig.~2 of \cite{bggs07} at the frequency of $325\,
\mathrm{MHz}$) consists of an inner region with slower apparent
drift velocity, surrounded by an outer region with larger
intensity of subpulses and steeper drift. Multifrequency
observations of \cite{bgg09} at $157, 244, 325, 610$ and $1060\,
\mathrm{MHz}$ show that at lower frequencies the subpulses become
weaker and may be seen only in the outer regions at $244\,
\mathrm{MHz}$ and only in the trailing outer region at $157\,
\mathrm{MHz}$. The observing geometry, \ie the values of the
inclination angle and of the impact angle, is not uniquely
determined for this pulsar and the polarization profile admits
several interpretations. However, based on the average
polarization behaviour,  \cite{qmlg95} concluded that the
inclination angle of PSR B0818-41 should be small. Unique nulling
properties of PSR B0818-41 are studied in \cite{bgg10}.

\subsubsection{Subpulse phenomenology}

The value of $P_3 = 18.3\pm 1.6 P$ was found in \cite{bggs07},
observing at the frequency $325\, \mathrm{MHz}$, by means of the
fluctuation spectrum analysis, and the same value was confirmed
later in \cite{bgg09} for all other observing frequencies. In the
inner part of the subpulse drift region, where several (typically
3 to 4) subpulse tracks are observed within one pulse, the value
of $P_2$ was found from the second peak of the autocorrelation
function to be ${17.5 ^\circ}  \pm {1.3^\circ }$
\citep{bggs07}. In the outer regions of the profile the value of
$P_2$ is larger (already from the visual inspection of the
subpulse tracks) and not easily measurable, because typically no
more than one subpulse per pulse is seen in these regions.
Estimations for the different observing frequencies and different
parts of the profile can be found in \cite{bgg09}, while the
average value of $P_2$ may be taken around $28^\circ$
\citep{bggs07}.

As indicated in \cite{ggks04}, the measured values of the periods
$P_2$ and $P_3$ are not necessarily equal to the true intrinsic
ones. The value of $P_3$ may be affected by aliasing, which starts
to play a role when $P_3<2P$. The value of $P_2$ is affected by
the finite time required for the line of sight to traverse the
polar cap as well as by the difference between the longitude $l$
along the pulse (in which we measure $P_2$) and the azimuthal
coordinate $\phi$ around the magnetic axis. However, if we assume
that there is no aliasing, the correction to the measured value of
$P_2$ due to the finite traverse time is given by a factor of
${1}/{[1+P_2/(360^\circ P_3)]}$ [derived from the equation
(5) of \cite{ggks04}], which in our case is $\sim 0.997$.
As long as we are
not concerned with the structure of the carousel as a
whole and we don't consider the possibility of aliasing in
our calculations, we assume everywhere that the measured
values of $P_2$ and $P_3$ are equal to the intrinsic
ones.

\subsubsection{Analysis of the subpulse drift}
As in the previous subsection, we start the analysis from
the determination of $\bar{r}$, corresponding to the
measured subpulse drift velocity of PSR B0818-41. Choosing
for this estimation $\chi = 0$ and $\xi = 0.5$ and using
the other known parameters of the pulsar, we get $\bar{r}_0=1.011$
for $\omega_{D} = - 0.956 ^\circ/P$ (this value of $\bar{r}_0$
depends very weakly on the chosen $\xi$ and $\chi$,
provided the inclination angle is small). For comparison,
assuming that the temperature of the polar cap is $T=3\times 10^6
\mathrm{K}$, the altitude of the PFF $h^{\rm HA}$ lies in the
range $0.013-0.106$ for the different values of
$f_{\rho}$ and $h^{\rm HM}=0.182$, while for the temperature
$T=4\times 10^6 \mathrm{K}$ the corresponding values for
the $h^{\rm HA}$ are $0.0097-0.0796$ with the same
$h^{\rm HM}$. So, the estimated value for the altitude lies
close to the lower boundary obtained through the PFF approach.

\begin{figure}
\begin{center}
\includegraphics[width=0.35\textwidth]{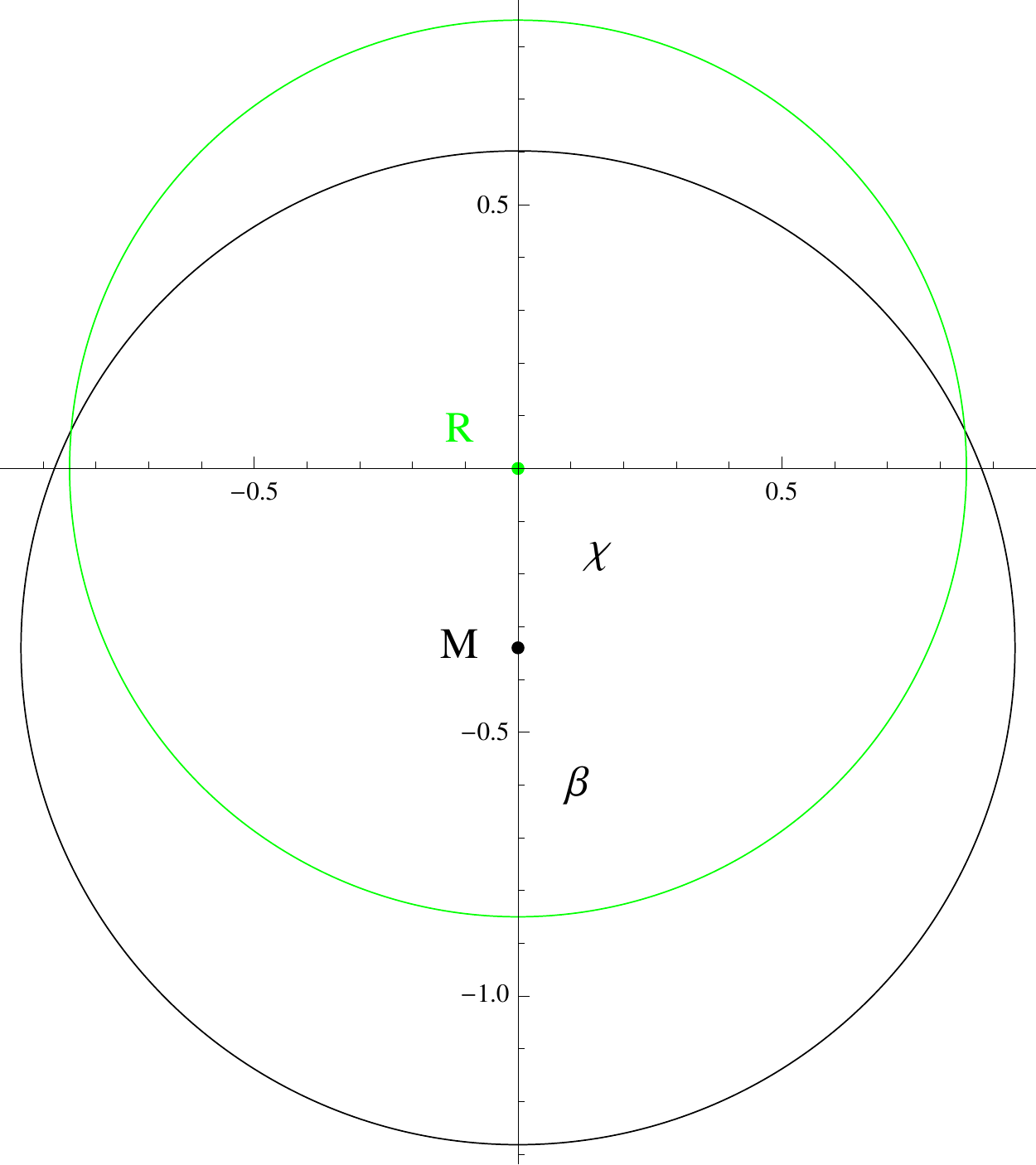}
\includegraphics[width=0.49\textwidth]{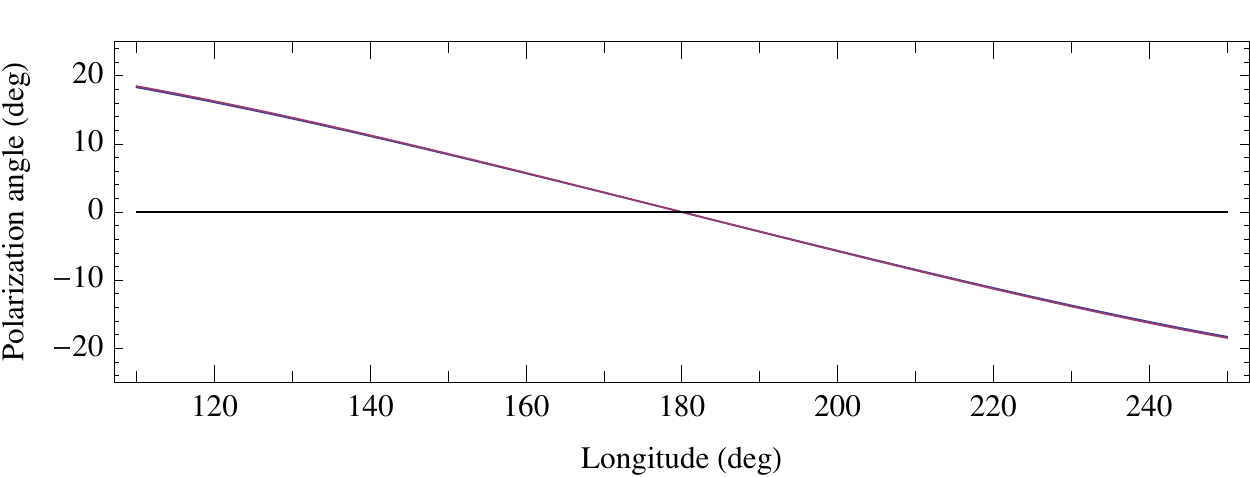}
\end{center}
\caption{Upper panel: observing
  geometry of PSR B0818-41 (the coordinate axes show the
  values of the angular coordinate $\theta$ in
  degrees). The black circle indicates the polar cap,
  while the green circle indicates the trajectory of the
  line of sight of the observer.
  Lower panel: Polarization angle as a function
  of longitude in our model.
}
\label{diagram}
\end{figure}

Fig.~\ref{diagram} is devoted to the illustration of the
geometry of PSR B0818-41,
with the inclination
angle between the magnetic and the rotational axes
$\chi=0.34^\circ$ and the impact angle
$\beta=0.51^\circ$. The considered geometry corresponds
to an outer line of sight, which, within the SCLF model,
 naturally results in negative value of the subpulse
 drift velocity.
The chosen inclination and impact angles are much smaller than
those  suggested before in the literature\footnote{ The two
geometries proposed by \cite{bgg09} have $\chi = 11^\circ\ ,\
\beta=-5.4^\circ $ (G-1) and $\chi = 175.4^\circ \ ,\
\beta=-6.9^\circ$ (G-2).}. These small values are required by the
fact that the size of the polar cap is very small
$\Theta=0.94^\circ$ at the considered altitude. However, one may
notice that the G-2 geometry of  \cite{bgg09}, which gives the
best fit to the polarization angle profile of PSR B0818-41,
effectively corresponds to an outer line of sight with $\chi
= 180^\circ - 175.4^\circ  = 4.6^\circ $ and $\beta = 6.9^\circ$,
so that $\sin\beta/\sin\chi = 1.5$. Based on this, we chose the
inclination and impact angles of our geometry to satisfy
$\beta/\chi=1.5$ in order to match the polarization profile of the
pulsar. The polarization angle, calculated through the ``Rotating
Vector Model'', is plotted in the lower panel of the
Fig.~\ref{diagram} as a function of the longitude and reproduces
well the observational data [\cf Fig.~6 of \cite{bgg09}].

\begin{figure}
\includegraphics[width=0.45\textwidth]{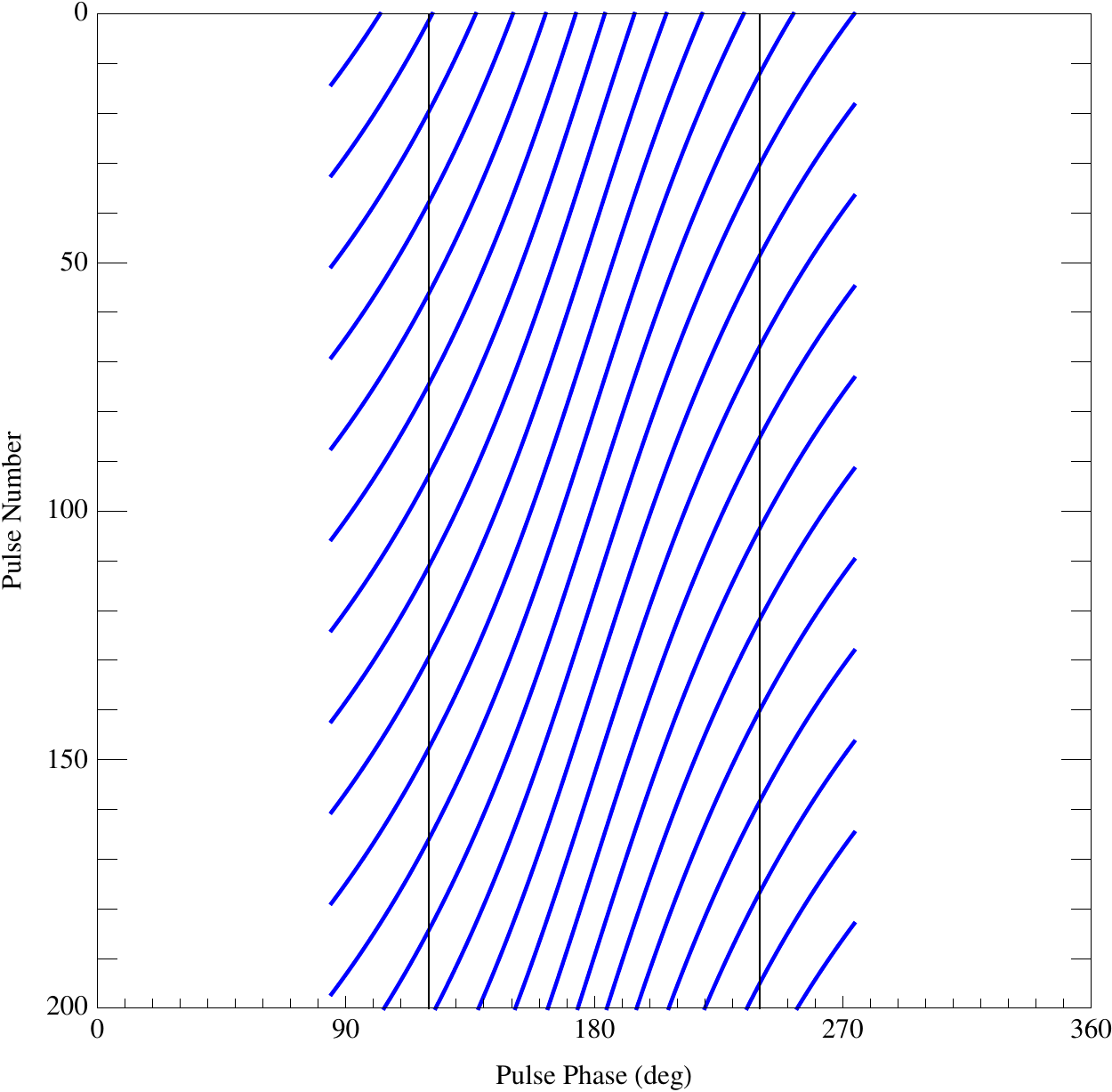}
\caption{Curved subpulse driftbands of PSR B0818-42
  generated through the SCLF model (solid blue lines), to
  be compared with those reported in Fig.~2 of \citet{bggs07}.
  Tracks are plotted for the whole longitude range of the average profile,
  while vertical black lines indicate the boundaries of the region where drifting subpulses are actually observed.} \label{tracks}
\end{figure}

We can use the geometry of Fig.~\ref{diagram} to model
the observed behavior of the subpulse tracks. Starting
from the trailing end of the profile (because the drift
velocity is negative) we evolve the azimuthal coordinate
$\phi$ with time using the projected velocity
(\ref{omegaproj}). At each time step we calculate the
corresponding pulse longitude using relation (\ref{l})
and add a new subpulse track every $P_3=18.3 P$. The
resulting pattern of the subpulse motion for $200$ pulses
as a function of pulse longitude is shown in
Fig.~\ref{tracks} through blue solid lines, which may be
directly compared to Fig.~2 of \cite{bggs07}. The solid
vertical lines indicate the region, where the subpulse
drift is actually observed. We estimated its width as
$120^\circ$, corresponding to the pulse longitudes
$140-260^\circ$ of Fig.~2 of \cite{bggs07}. Note that the
center of the profile there is slightly shifted with respect
to $180^\circ$. 
When compared to the observational data, we
find that the SCLF model is able to reproduce the curved
subpulse tracks reasonably well.

The pattern of the subpulse tracks obtained with our
model is symmetric by construction.
However, according to the observations of \cite{bgg09},
the drift bands of the trailing outer region of the pulse
appear to be steeper than the drift bands of the leading
outer region.
Although a clean explanation for this effect is still
missing, we argue that some degree of asymmetry can be due
to the effects of retardation, aberration and
refraction of the signal in the outer magnetosphere
\citep{gg01,gg03,p00,wshe03}.

We emphasize that the original spark model of \cite{gs00}, on the
basis of very general arguments, predicts that the pulsar polar
cap should be densely filled by equidistant equal-size sparks. On
the contrary, the carousel pattern proposed later in
\cite{eamn12}, \cite{bggs07}, \cite{bgg09} has larger and wider
separated sparks in the outer ring in order to explain the
observed subpulse behavior (compare Fig.~1 of \cite{gs00} with
Fig.~10 of \cite{eetal05}). We find it interesting that, within
the SCLF model, it is possible to  explain the observed curved
subpulse tracks by means of the velocity variations only, without
breaking the assumption that the features responsible for the
subpulses are equal in size and equidistant. This supports the
argument,  already presented in the previous subsection, according
to which the variations of $P_2$ with the pulse longitude,
observed for many pulsars, may be completely explained by the
variability of the subpulse drift velocity across the polar cap,
while the value of $P_3$, which seems to be independent of the
pulse longitude and even of the observing frequency,  should
reflect an intrinsic characteristic property  of the
individual pulsar.

One interesting observation made in \cite{bgg09} is that the
leading and trailing outer regions of the pulse profile maintain a
unique phase relationship, with the maximum of the energy in the
trailing component being shifted in time by $\sim 9 P$ with
respect to the maximum of the energy in the leading component. Based
on this observation, the authors propose an elegant solution to
the aliasing problem, arguing that the considered shift may not be
explained without aliasing and suggesting the model of $20$ sparks
ring with first-order alias and a true drift velocity of $19.05^\circ / P$ 
as the most plausible description of the system. As an
alternative, we propose that the position
  of the picks of the pulse profile is not strictly
  determined by the position of the sparks in the outer
  ring,
  but modulated by the outer regions of the
  magnetosphere. One may assume, as it is customary, that the major
  radio emission mechanism of the pulsar is due
  the formation of the
  secondary plasma from the energetic photons emitted by
  the primary particles. Near the axis this process is
  negligible due to the large curvature radius of the
  magnetic field lines, while on the edges of the polar
  cap region the acceleration potential itself drops to
  zero. This produces the ``hollow cone'' distribution
  of the secondary plasma in the magnetosphere above the
  PFF (see \cite{p00}, \cite{wshe03}, where this model is
  used for the study of magnetospheric refraction, also \cite{fkk06}). We
  therefore believe
  that such a distribution modulates the emission
  profile and changes the position of the maxima with
  respect to the position of the outer ring of
  sparks. However, this issue is beyond the scope of
  the present paper and we leave it for a future study.

%%%%%%%%%%%%%%%%%%%%%%%%%%%%%%%%%%%%%%%%%%%%%%%%%
\section{Conclusions}
\label{conclusions}
%%%%%%%%%%%%%%%%%%%%%%%%%%%%%%%%%%%%%%%%%%%%%%%%%

The phenomena of drifting subpulses are typically
explained by resorting to the vacuum or to the partially
screened gap models of pulsar magnetospheres.
For example, the partially screened gap model allows for the formation of
a spark carousel due to discharges of the large potential
drop through the inner polar gap and these sparks are
thought to be responsible for the appearance of
subpulses.
%At the same moment the subpulse drift
%velocities this model predicts are smaller than those of
%the pure vacuum gap model and comparable with the
%observed values.
However, it has  been recently shown by
\citet{lt12} that the expression used for the  estimation of
the subpulse drift velocities, both in the vacuum and
in the partially screened gap model, is not accurate enough.

On the other hand, considering the pulsar magnetosphere as a
global object, one can propose alternative mechanisms for the
formation of distinct emitting features representing subpulses and
in this paper we have reconsidered the ability of the
Space-Charge Limited Flow model to explain this phenomenology.
The SCLF model provides analytical solutions for the scalar
potential in the polar cap region of the pulsar magnetosphere in
case of free outflow of the charged particles from the surface of
the star. Hence, the drift velocity of subpulses along the pulse
can be interpreted in terms of the plasma drift velocity, which in
turn depends on the gradient of the scalar potential rather
than on its absolute value.

After considering a selected sample of sources taken from
the catalog of \cite{wes06}, we have found the
following conclusions:

\begin{itemize}
\item the SCLF model predicts the subpulse drift
  velocities compatible to the observed ones at heights
  above the surface of the star close to the pair
  formation front;

\item the angular dependence of the plasma drift velocity in
  the SCLF model provides a natural explanation for the
  variation of the subpulse separation $P_2$ along the
  pulse. In particular it may explain the curved subpulse
  driftbands of PSR B0818-41 and the range of the observed
  subpulse velocities of PSR B0826-34.

\end{itemize}

These results suggest that the role of the SCLF
model in explaining the drifting subpulse phenomena has been
underestimated, calling for additional investigations and
systematic comparisons with all available observations.

%%%%%%%%%%%%%%%%%%%%%%%%%%%%%%%%%%%%%%%%%%%%%%%%%%%%%%%%%%%%%%%%%%%%%%%%%
\section*{Acknowledgments}
%%%%%%%%%%%%%%%%%%%%%%%%%%%%%%%%%%%%%%%%%%%%%%%%%%%%%%%%%%%%%%%%%%%%%%%%%

This research was partially supported by the Volkswagen Stiftung
(Grant 86 866). We would like to thank Luciano Rezzolla for
support, as well as Farruh Atamurotov for important discussions. V.M. 
thanks Christian Ott for hospitality at Caltech.
B.A.'s research is supported in part by
the project F2-FA-F113 of the UzAS and by the ICTP through the
OEA-NET-76, OEA-PRJ-29 projects.
O.Z. acknowledges support to the European Research Council under
the European Union's Seventh Framework Programme (FP7/2007-2013)
in the frame of the research project \textit{STiMulUs}, ERC Grant
agreement no. 278267. We would like to thank 
our referee for very valuable comments, which
helped us to improve the quality of the manuscript.

\bibliography{subpulse_SCLF}{}
\bibliographystyle{mn2e}

%%%%%%%%%%%%%%%%%%%%%%%%%%%%%%%%%%%%%%%%%%%%%%%%%%%%

\renewcommand \theequation {A.\arabic {equation}}
\setcounter{equation}{0}

\appendix
\section{Pair formation front}
\label{Ap1}

All mechanisms proposed for the generation of radio emission in
the pulsar magnetosphere require the presence of an
electron-positron plasma. Within the SCLF model, primary
particles, extracted from the surface of the star by the
rotationally induced electric field, accelerate in the inner
magnetosphere and emit high energy photons, which in turn produce
electron-positron pairs in the background magnetic field. The
three main processes responsible for the emission of photons by
the primary particles are curvature radiation (CR), nonresonant
inverse Compton scattering (NRICS) and resonant inverse Compton
scattering (RICS). Copious pair production in the open field lines
region leads to the screening of the accelerating electric field
and stops the acceleration of the particles above the so-called
pair formation front (PFF).

The determination and even the definition of the PFF is not a
trivial task. It was thoroughly investigated in the framework of
the SCLF model by \citet{ha01} and \citet{hm01,hm02}, with
slightly different approaches. \citet{ha01} define the location of
the PFF as the place  where the number of pairs created per
primary particle is equal to $\kappa$, which means that the space
charge density is large enough to screen the accelerating
component of the electric field. \citet{hm01,hm02}, instead,
locate the PFF front where the first electron-positron pair is
produced. In \citet{hm01,hm02} it is shown that the full screening
of the accelerating electric field is not even possible for many
pulsars, while the pair formation front is still formed. The
difference between these two approaches affects significantly the
inverse Compton scattering, while the results for the curvature
radiation are essentially the same.

The expressions for the height of the PFF in units of the stellar
radius obtained in \cite{ha01} are

\begin{eqnarray}
\label{HA}
h_{\rm CR}^{\rm HA} &=& 0.678 B_{12}^{-5/6} P^{19/12} f_{\rho}^{1/2} \ ,  \\
h_{\rm NRICS}^{\rm HA} &=& 0.119 B_{12}^{-1/2} P^{1/4} T_6^{-1} f_{\rho}^{1/2} \ ,  \\
h_{RICS}^{HA} &=& 12.0 B_{12}^{-7/3} T_6^{-2/3} f_{\rho} \ .
\end{eqnarray}
Here $B_{12}=B/10^{12}$ and $T_{6}=T/10^6$. The
quantity $f_{\rho}$, which describes the curvature of the field
lines in the considered regions, changes from $f_{\rho} = 0.011
P^{-1/2}$ for the multipolar field with radius of curvature equal
to the stellar radius, to $f_{\rho} = 1$ for the dipolar field. In
Fig.~\ref{fig1} these two cases correspond to the lower and upper
boundaries of the blue shaded regions.\footnote{Here for NRICS we
report only the values obtained in the Klein--Nishina regime, as it
will dominate for typical pulsar parameters.}

The expressions for the height of the PFF in units of the stellar
radius obtained in \cite{hm01,hm02} are

\begin{eqnarray}
\label{HM} h_{\rm CR}^{\rm HM}&\approx& 0.03 \left\{
    \begin{array}{@{} l c @{}}
      1.9 P_{0.1}^{11/14} B_{12}^{-4/7}  \text{if} \,\,\,P_{0.1}^{9/4}<0.5 B_{12}\ , \\
      3.0 P_{0.1}^{7/4} B_{12}^{-1}  \text{if} \,\,\,P_{0.1}^{9/4}>0.4 B_{12}
    \end{array}\right.  \\
h_{\rm NRICS}^{\rm HM}&\approx& 0.01 \left\{
    \begin{array}{@{} l c @{}}
      3 (P/B_{12})^{2/3}  \text{if} \,\,\,P\lesssim 0.4 B_{12}^{4/7}\ , \\
      4 P^{5/4}/B_{12}  \text{if} \,\,\, P\gtrsim 0.4 B_{12}^{4/7}
    \end{array}\right.  \\
h_{\rm RICS}^{\rm HM}&\approx& 0.01 \left\{
    \begin{array}{@{} l c @{}}
      7 P^{2/3} B_{12}^{-1}  \text{if} \,\,\,P\lesssim 0.1 B_{12}^{6/7}\ . \\
      17 P^{5/4} B_{12}^{-3/2}  \text{if} \,\,\,P\gtrsim 0.1 B_{12}^{6/7}
    \end{array}\right.
\end{eqnarray}

The values for the PFF shown in Fig.~\ref{fig1} are $\min(h_{\rm
CR}^{\rm HM},h_{\rm NRICS}^{\rm HM},h_{\rm
  RICS}^{\rm HM})$ for the blue points, $\min(h_{\rm
  CR}^{\rm HA},h_{\rm NRICS}^{\rm HA},h_{\rm RICS}^{\rm
  HA})$ with $f_{\rho} = 0.011 P^{-1/2}$ for the lower
boundaries of the blue shaded regions and $\min(h_{\rm
  CR}^{\rm HA},h_{\rm NRICS}^{\rm HA},h_{\rm RICS}^{\rm HA})$ with $f_{\rho} = 1$ for the upper boundaries of the blue shaded regions.

\end{document}